%% file: bct_Feb19_2018.tex
\documentclass[11pt, oneside]{amsart}
\pdfoutput=1
\usepackage{geometry}
\usepackage{lmodern}
\usepackage{amsmath}
\usepackage{amssymb}
\usepackage{amsthm}
\usepackage{comment}
\usepackage{color}
\usepackage[usenames,dvipsnames]{xcolor}
\usepackage{graphicx}
\usepackage{subfig}
\usepackage{soul}
\usepackage[countmax]{subfloat}
\usepackage{float}
\usepackage{tikz}
\usepackage{enumerate}
\usepackage{rotfloat}
\usepackage{setspace}
\usepackage{esint}
\usepackage{lscape}
\usepackage[authoryear]{natbib}
\definecolor{MyDarkBlue}{rgb}{0,0.08,0.45}
\usepackage[unicode=true,pdfusetitle, bookmarks=true,bookmarksnumbered=false,bookmarksopen=false, breaklinks=false,pdfborder={0 0 1},backref=section,colorlinks=true, linkcolor = MyDarkBlue, citecolor = MyDarkBlue]{hyperref}
\usepackage{breakurl}
\usepackage[para]{threeparttable}
\allowdisplaybreaks

\PassOptionsToPackage{normalem}{ulem}
\usepackage{ulem}

\makeatletter

\hypersetup{colorlinks=true, linkcolor=MyDarkBlue}

\newtheorem{proposition}{{\bf\sc Proposition}}

\DeclareMathOperator*{\supp}{supp}

\providecommand{\g}{{\bf g}}

\makeatother

\title[Investigating Network Knowledge]{Seeing the forest for the trees? An investigation of network knowledge}

\author{Emily Breza$^{\dagger}$ }
\author{Arun G. Chandrasekhar$^{\ddagger}$}
\author{Alireza Tahbaz-Salehi$^{\star}$ }
\date{This Version: \today}

\thanks{We thank Abhijit Banerjee, Joshua Blumenstock, Steven Durlauf, Paul Goldsmith-Pinkham, Ben Golub, Matt Jackson, Cynthia Kinnan, Pooya Molavi, Melanie Morten, Suresh Naidu, Xu Tan, and Leeat Yariv for helpful comments. Financial support from the NSF under grants SES-1156182 and SES-1326661 is gratefully acknowledged. We thank Shobha Dundi, Devika Lakhote, Tithee Mukhopadhyay, Gowri Nagraj, and Ambika Sharma for excellent research assistance.
}
\thanks{$^{\dagger}$Department of Economics, Harvard University}
\thanks{$^{\ddagger}$Department of Economics, Stanford University}
\thanks{$^{\star}$Kellogg School of Management, Northwestern University}

\begin{document}

\maketitle
\begin{abstract}


This paper assesses the empirical content of one of the most prevalent assumptions in the economics of networks literature, namely the assumption that decision makers have full knowledge about the networks they interact on. Using network data from 75 villages, we ask 4,554 individuals to assess whether five randomly chosen pairs of households in their village are linked through financial, social, and informational relationships. We find that network knowledge is low and highly localized, declining steeply with the pair's network distance to the respondent. 46\% of respondents are not even able to offer a guess about the status of a potential link between a given pair of individuals. Even when willing to offer a guess, respondents can only correctly identify the links 37\% of the time. We also find that a one-step increase in the social distance to the pair corresponds to a 10pp increase in the probability of misidentifying the link. We then investigate the theoretical implications of this assumption by showing that the predictions of various models change substantially if agents behave under the more realistic assumption of incomplete knowledge about the network. Taken together, our results suggest that the assumption of full network knowledge (i) may serve as a poor approximation to the real world and (ii) is not innocuous:  allowing for incomplete network knowledge may have first-order implications for a range of qualitative and quantitative results in various contexts.\\


\end{abstract}

\setcounter{page}{0}

\newpage

\section{Introduction}

One of the most prevalent assumptions in the network economics literature is that decision makers have correct, complete, and common knowledge about the structure of the networks they interact on. For instance,  various models of social learning over networks \citep*{banerjee1992simple, SmithSorensen, mossel2015strategic}, the literature on network games of strategic complementarities \citep*{ballester2006s, CalvoREStud, Ozan}, as well as the growing literature on the econometrics of network formation \citep*{sheng2016structural,leung2014random,menzel2015strategic,de2017identifying} implicitly or explicitly assume that agents have full knowledge about the underlying network structure.

In this paper, we assess the empirical content and theoretical implications of this assumption. Using observational data, we document substantial departures from full network knowledge, underscoring that the assumption of complete network information may serve as a poor approximation to the real world. We then investigate the theoretical implications of relaxing this assumption in each of the three different domains mentioned above --- social learning, network games, and estimation of models of network formation. Our results illustrate that, while a sensible first step, this assumption is not innocuous: violations of full network knowledge may have qualitatively and quantitatively important implications for the predictions of each of these models.

Our empirical investigation uses data collected in 75 villages in Karnataka, India, where we have previously collected detailed network data across all 16,500+ households \citep{gossip2016}. Our network data consists of information on whether or not a link exists across a number of informational, financial, and social dimensions for over 98\% of pairs of households in each village. Against this backdrop, we returned to the 75 villages and asked 4,554 respondents about existence of various forms of linkages between five pairs of individuals from their village. In particular, we first asked the respondents whether they are unable to offer a response --- either because they are not certain enough or because they do not know at least one of the individuals in the corresponding pair. If the respondents were able to offer an answer, we then solicited their guess about the link's existence. 

As our main empirical finding, we document a substantial lack of knowledge about the network structure, both in terms of respondents' uncertainty as well as the extent to which they correctly identified the presence or absence of links in the underlying network. 

Consistent with a basic lack of network knowledge, we find that almost half (46\%)  of respondents reported that they ``don't know'' the link status between a given pair $(j,k)$ of individuals in their village. We then regressed the dummy variable capturing respondent's uncertainty regarding the link on the respondent's distance to $j$ and $k$, controlling for the respondent's eigenvector centrality, the average centrality of $j$ and $k$, and a number of demographic controls (such as caste, amenities, geography). We find that respondents are more likely to express uncertainty about the existence of a link between $j$ and $k$ the further they are in the network: a one-step increase in the average distance of $j$ and $k$ to the respondent is associated with a 12.2 percentage point (pp) increase ($p$=0) in the probability of expressing uncertainty. This effect is sizable, specially relative to the mean of 46\%.

We then investigate the quality of the respondents' information by regressing whether the respondent correctly identifies the link status of a given pair $(j,k)$ on network distance, while controlling for centralities and demographic covariates. We find that the quality of the respondent's information also declines steeply in social distance: a one-step increase in the average distance of $j$ and $k$ to the respondent is associated with a 9.3pp decline ($p$=0) in the respondent's probability of correctly identifying that $j$ and $k$ are linked in the underlying network. This also captures a sizable effect, especially given that the mean rate of correctly identifying a link's status is only 20\%.\footnote{This indicates systematic misspecification: respondents are more likely than not to think a link exists. A back-of-the envelope calculation shows that they must be overestimating the average degree by at least 38\%.} Even conditional on expressing certainty, we find that being one step further is associated with a 1.1pp  to 2.1pp decline ($p$=0.005) in the probability of correctly identifying that $j$ and $k$ are  linked.

The bulk of the aforementioned results are not explained by other demographic covariates. We include geographic distance between respondents and $jk$ pairs, dummies for whether the respondent is of the same caste or subcaste, has the same household amenities (e.g., electrification, roof type, ownership) as $j$ and $k$, interactions of all of these covariates with actual $jk$ link status, as well as village fixed effects and even respondent fixed effects. Across specifications, our conclusions are predominantly unchanged both qualitatively and quantitatively. We then conduct a post-LASSO exercise to determine a sparse set of covariates that predict our key outcomes, namely, the respondent's correct assessment of link status as well as her claim whether to know or not know the status of a given link. Both in terms of correct assessments and degree of certainty, we find that it is the average network distance to $j$ and $k$ that is selected (along with respondent centrality and the average centrality of $j$ and $k$). In contrast, except for caste and subcaste, all other demographic variables  --- including geographic distance between the respondent and $jk$ --- are not selected. This finding is consistent with a story where agents learn about the structure of the network through interactions on and through the network, where knowledge tends to be localized.

We next turn to the theoretical implications of our empirical findings by showing that relaxing the widespread assumption that the network structure is commonly known can have first-order implications for a range of qualitative and quantitative results in various contexts. As a first application, we focus on a simple model of Bayesian learning over social networks. We show that partial network knowledge may lead to identification problems that can impede learning, even though information would have been efficiently aggregated had individuals faced no uncertainty about the network structure. This rests on the observation that knowledge about the intricate details of the network structure provides the agents with valuable information on how to discern the correlations and redundancies in their neighbors' estimates. Our result thus demonstrates that predictions about learning dynamics may be sensitive to the extent and nature of agents' knowledge about the social network structure. 

As a second case study, we illustrate how assuming full network knowledge may lead to biased structural estimates if in fact decision makers face some uncertainty about the underlying network. We focus on the canonical network interaction game of \cite{ballester2006s} with strategic complementarities. This game, which serves as one of the workhorse models for studying strategic network interactions in the literature, allows for simple structural estimates, counterfactuals, and policy prescriptions. We show that uncertainty about the extent of strategic complementarities outside one's neighborhood results in a systematic shift in equilibrium actions (relative to the complete information benchmark). Hence, ignoring the possibility of incomplete network knowledge may lead to biased structural estimates for the degree of complementarities, mis-specified cost-benefit analyses, and potentially counterproductive intervention policies (such as identifying ``key players'' to target for interventions).

Finally, we shift our attention to investigating the implications of network knowledge for parameter identification in network formation games. We study a fairly standard econometric model of network formation, where the literature has grappled with partial identification of parameters  when agents have complete information about the realization of pairwise preference shocks. However, we show that the equilibrium network in the game with incomplete information provides the econometrician with sufficiently rich observations to point-identify all structural parameters. This result suggests that introducing the (more realistic) assumption of incomplete information may transform a model that is fundamentally hard to identify to a straightforwardly estimable model. 

Our work is not the first paper to introduce incomplete information into the above mentioned contexts. For instance, in the context of social learning, \citet{acemoglu2011bayesian} and \citet*{LobelSadlerTE} allow for randomly generated  networks of various kinds. Similarly, papers such as \citet*{leung2015two} and more recently \citet*{RidderSheng} study network formation  with incomplete information. Our goal here is primarily pedagogical, strongly motivated by the data. Specifically, our results are meant to clarify the extent to which a potentially unrealistic assumption plays a critical role in a wide variety of applications and how incorporating partial or lack of network knowledge into fairly standard models can lead to vastly different conclusions. 

In addition to the network theory and econometrics literatures discussed above, our paper is related to works in sociology that explore limits to knowledge about others' traits as well as perceptions of friendships in networks \citep{friedkin1983horizons,krackhardt1987cognitive,krackhardt2014preliminary}. For instance, using network data among several departments at the University of Chicago and Columbia University, \cite{friedkin1983horizons} showed that a respondent $i$ was less likely to know about another faculty member $j$'s current research if the respondent was further from $j$ in the network. This is in line with findings of \citet{alatas2012network}, who show that subjects are much less likely to know about the wealth status of individuals in their village who are further away in the network. Relative to these findings, our paper emphasizes that knowledge about the network itself is similarly limited and localized through the network.\footnote{\citet*{krackhardt1987cognitive,krackhardt2014preliminary} was the first to advocate for collecting data not only about network interactions, but also about others' perceptions of such interactions. \citet*{krackhardt1987cognitive} called these cognitive social structures (CSS). Our data could be interpreted as sampling from CSSs in 75 distinct networks.}

The rest of the paper is organized as follows. Section \ref{sec:setting} describes the setting, data, and sample statistics. Section \ref{sec:network_distribution} contains our main empirical findings, where we document how network knowledge relates to the average distance between the respondents and the pair of individuals they are being inquired about. In Section \ref{sec:importance}, we discuss how our findings regarding limited network knowledge can be relevant for theoretical, applied, and econometric work. Section \ref{sec:conclusion} is a conclusion. All proofs are provided in Appendix \ref{sec:proof}. An online appendix contains some additional empirical results.

\section{Setting and Data}\label{sec:setting}

\subsection{Network data collection} We collected data in 75 villages in Karnataka, India, where we have previously worked. The villages span 5 districts around Bangalore where we have collected network data in the past: Wave I in 2006 \citep{banerjee2013diffusion} and Wave II in 2012  \citep{gossip2016}. In 2012, we collected network data from 89\% of the over 16,500 households across the 75 villages. These links spanned financial, social, and informational relationships across 12 dimensions.\footnote{(1) whose house the respondent visits, (2) who visits her house, (3) kin, (4) whom they socialize with, (5) who gives information when there is a medical need, (6) whom they gives advice to, (7) whom they get advice from, (8) whom they lend material goods to, (9) whom they borrow material goods from, (10) whom they borrow money from, (11) whom they lend money to, (12) with whom they go to pray (e.g., at a temple or mosque).} Because such a high share of households were surveyed and every household could name links to any other household in the census in each village, we have 98.8\% of all links in the resulting undirected, unweighted network. It is against this backdrop that we conducted a subsequent survey to explore network knowledge.

\subsection{Knowledge data collection} We conducted a network knowledge questionnaire in 2015 by asking 4,554 randomly selected  individuals  across the 75 villages about network relationships between various other individuals in their community. We asked every respondent $i$ about network relationships among five distinct pairs $j$ and $k$. This was stratified in the following way. We ensured that for each $i$, one $jk$ pair of each distance 1--1.5, distance 2--2.5, distance 3--3.5 distance 4--4.5, and distance 5+ (including unreachables) was selected in the survey, using our pre-existing network data.\footnote{We also asked $i$ about several demographic traits of $j$ and $k$. These include if they have children, household size, main household occupation, monthly income, television ownership, religious details, political disposition, and land ownership.} For each $jk$ pair we asked about the existence of specific types of network relationships: informational, financial, and social. We analyze responses to informational, financial, and social relationships, denoted by index $r$.\footnote{Specifically, we collapsed some of the \cite{banerjee2013diffusion} survey, leaving us with 3 dimensions: (1) social interaction; (2) borrowing/lending of money; (3) giving/receiving advice before an important decision;}
  The resulting object is a graph $\g_r$ for each $r$.

For every pair $jk$ and relationship type $r$, we asked  $i$ whether $g_{jk,r}=1$ or $g_{jk,r}=0$. The respondent could tell us their estimate or tell us that they ``Don't know.'' A response of ``Don't know'' could arise for two reasons: either the respondent does not feel certain enough to offer a guess or because the respondent does not know who either $j$ or $k$ are. 
 The respondent $i$'s response is therefore $h_{i,jk,r} \in \{1,0,\text{Don't Know}\}$.\footnote{Note that even though the subjects report whether or not they believe a given pair of individuals are linked, we do not observe the respondents' subjective beliefs (i.e., the probability that $i$ assigns to the existence of a link $j$ and $k$).} 
 
 Our outcomes of interest are whether the respondent correctly identified link status, i.e.,
\begin{align*}
y_{i,jk,r} & = 1\{h_{i,jk,r} = g_{jk,r}\},
\end{align*} 
and whether they were sure enough to offer a guess to being with, i.e.,
\begin{align*}
\text{DK}_{i,jk,r} & = 1\{h_{i,jk,r} = \text{Don't Know}\}.
\end{align*}
We are chiefly interested in how $y_{i,jk,r}$ and $\text{DK}_{i,jk,r}$ depend on the network distance between the respondent $i$ and the pair $jk$, conditional on the centralities of the nodes involved and demographic covariates.  While we consider the respondents' outcomes $y_{i,jk,r} = 1\{h_{i,jk,r} = g_{jk,r}\}$ for each type of relationship $r$ separately in our regressions, we follow previous work and measure distances and centralities in the union network $\g = \cup_r \g_r$, according to which two agents are assumed to be linked if either agent reports having any of the 12 relationship dimensions with the other.\footnote{This choice is made to generate the regressors as independent variables because we believe that this generates the most natural definition of social distance for our setting. Individuals might learn about a financial relationships existing in other parts of the network, for example, through financial ties but also through information and social ties.}

\subsection{Sample statistics}

Respondent characteristics are summarized in Table \ref{tab:summary}, Panel A. 59\% of the respondents were female and the average age in the sample was 42.55. Males were typically born in the village (51\%) whereas most females married in (only 11\% were born in the village). The average degree is 19.5 which indicates a sparse network (the average number of households in the village is 196). Panel B describe pair characteristics. 99\% of the pairs belonged to the same connected component as the respondent (i.e., the pair was `reachable'), and in our stratified sample, 34\% of the time they were linked. On average, the distance from $i$ to $jk$ was 2.13, and 59\% of the time they were of the same caste category. Panel C presents our knowledge survey outcomes. 46\% of the time the respondents expressed that they did not know the linking status of the pairs in question. When willing to offer a guess, the respondents correctly identified the linking status of the pair only 20\% of the time. Conditional on a link existing, a correct guess was offered 57\% of the time. This immediately illustrates that respondents systematically overestimated the existence of a link.\footnote{Assuming that each respondent simply offers a response about the existence of a link uniformly at random, a back-of-the-envelope calculation shows that the implied density of the network by our respondents is at least 37.9\% higher than that in the data.}

\

\section{Network Knowledge in the Network}\label{sec:network_distribution}

\subsection{Knowledge and distance}

Our summary statistics in Table \ref{tab:summary} already indicate poor average knowledge about the network.  We next explore how the relative position of nodes influences $i$'s ability to know whether $j$ and $k$ are linked. 

To this end, we run core regressions of the following form:
\begin{align}
Y_{i,jk,r} & =\alpha+\theta g_{jk,r}+\lambda_r +\beta^{\text{not}}\cdot \left(\frac{\text{dist}\left(i,j\right)+\text{dist}\left(i,k\right)}{2}\right)\cdot (1-g_{jk,r})  \label{eqn:main}\\
&+\beta^{\text{link}}\cdot\left(\frac{\text{dist}\left(i,j\right)+\text{dist}\left(i,k\right)}{2}\right) \cdot g_{jk,r} \nonumber \\
   & +\delta^{\text{not}}\cdot\text{Avg. Centrality}_{jk}\cdot (1-g_{jk,r})+\delta^{\text{link}}\cdot\text{Avg. Centrality}_{jk}\cdot g_{jk,r} \nonumber \\
 & +\gamma^{\text{not}}\cdot\text{Centrality}_{i}\cdot (1-g_{jk,r})+\gamma^{\text{link}}\cdot\text{Centrality}_{i}\cdot g_{jk,r} \nonumber \\
 & + X_{i,jk}'\eta+\epsilon_{i,jk,r}. \nonumber
\end{align}
Here, $Y_{i,jk,r}$ will either be $y_{i,jk,r}$, which is a dummy that measures whether $i$ correctly identified the link status between $j$ and $k$, or $\text{DK}_{i,jk}$, which is whether $i$ reported not knowing the status. 

Taking for now the case where $y_{i,jk,r}$ is the outcome variable of interest, coefficient $\theta$ measures the difference in the probability of being correct if the link $g_{ij,r}$ exists, whereas coefficients $\lambda_r$ are relationship type-fixed effects. Coefficient $\beta^{\text{not}}$ measures the marginal change in $i$'s probability of correctly identifying that $j$ and $k$ are not linked when the average distance increases by 1, while $\beta^{\text{link}}$ records how knowledge changes with distance when $g_{jk,r}=1$.\footnote{Note that even if respondents have biased guesses about whether links exist when they do not truly know (meaning that their guesses may differ on average from the true rate)  this does not affect the test of whether $\beta^{\text{not}}$ or $\beta^{\text{link}}$ is non-zero.} Similarly, $\gamma^{\text{not}}$ and $\gamma^{\text{link}}$ measure the marginal effect of being one standard deviation more central on the probability of being correct when the link does not exist, and that probability when the link does exist. $\delta^{\text{not}}$ and $\delta^{\text{link}}$ play the same role but now looking at $j$ and $k$'s centralities. Finally, $X_{i,jk}$ will include a vector of covariates: average geographic distance of $i$ from $j$ and $k$, their geographic centralities, and dummies for each of whether $i$ is of the same caste, subcaste, has the same electrification status, has the same roof type, and has the same ownership status for each of $j$ and $k$, as well as all these variables' interactions with linked status $g_{jk,r}$.

The main results can be seen in the raw data in Panels A--C of Figure \ref{fig:raw}. Panel A depicts the share of correct guesses by average network distance between $i$ and the $(j,k)$ pair. It is immediate to see that the fraction of correct guesses steadily declines in distance down to the unreachable pairs. Panel B illustrates that the share of respondents who do not know $jk$'s link status increases across network distance. In Panel C, we condition on having a view on the link status and then look at whether the guess is correct or not. We see in the raw data that there is a negative correlation between the share of correct guesses and the network distance, meaning that even beyond having a view on the link, there is residual information which is more accurate for pairs that are on average socially closer to the respondents. 

We can also consider how network knowledge varies with the centralities of $j$ and $k$. In Panels D--F of Figure \ref{fig:raw}, we plot the same outcomes of interest, varying average centralities.  Here, we see that when $j$ and $k$ are more central, $i$ is more likely to offer a correct guess, is less likely to say that she ``Doesn't know'' the link status, and has better information about the existence of a link conditional on having a view. Note that more central $j$ and $k$ are mechanically closer to any arbitrary $i$, so the relationship between centrality and knowledge is likely a direct consequence of the relationships in Panels A--C.\footnote{This finding may also help explain why respondents can identify the more central individuals fairly accurately \citep{gossip2016}.}

\subsubsection{How the likelihood of guessing correctly varies by distance}

Table \ref{tab:main} contains our main network distance results.  Panel A looks at how whether $i$ knows $jk$'s linking status, $y_{i,jk,r}$ depends on the relative network distances between parties and their centralities. Column 1 presents the regression just with the network position variables, column 2 adds the aforementioned vector of demographic controls contained in $X_{i,jk}$, column 3 adds village fixed effects, while column 4 also includes respondent fixed effects. Note that the specification with respondent fixed effects holds the demographic and network characteristics of the respondent fixed, comparing close versus far $jk$, within respondent.\footnote{Online Appendix \ref{sec:w_cent} presents the same tables with all network covariates.}

We focus on column 1 for exposition. We see that having $j$ and $k$ be on the same connected component as $i$ leads to a 6.5pp increase ($p$=0.001) in the probability of guessing the status correctly when there is no link. Further, this corresponds to a 54.7pp increase ($p$=0) in the probability of guessing the status correctly when there is a link.

Next, we study how respondents' knowledge depends on their network distance to the pair in question. Note that to understand the effect of distance, we need to condition on reachability through the network, as unreachable individuals have infinite distance to the respondent. We find that being one step further than $j$ and $k$ on average leads to a 1.8pp decline ($p$=0) in the probability of guessing correctly when there is no link, and more importantly a 10.9pp decline ($p$=0) in the probability of guessing correctly when there is a link. This is a very large effect relative to the mean (19.9\%). Furthermore, columns 2--4 illustrate that this pattern is robust to a wide range of specifications, clearly indicating that the quality of respondent's information is steeply declining in her network distance to the pair and is not explained by similarities in demographics such as wealth, caste, electrification, geography, and so on.\footnote{We note that adding controls does reduce the magnitude of the coefficient on reachability when there is a link. The other three coefficients are stable across specifications.}

\subsubsection{How the likelihood of being certain varies by distance}

Note that the correct guess outcome in Panel A combines both having a view on a given $jk$ relationship and guessing correctly about the existence (or not) of that relationship.  Panel B of Table  \ref{tab:main} focuses on the first component.  The specifications and control sets are the same as in Panel A, but now our outcome variable is $\text{DK}_{i,jk,r}$, a dummy for whether the respondent declared that they ``Don't know'' whether $j$ and $k$ are linked. 

We find that respondents are considerably more likely to have a view about the network structure local to them and are much more uncertain at larger distances. Once again, we focus on column 1 for exposition, though the results are largely robust to controls and fixed effects. Relative to a mean of 46\%, we observe that being reachable leads to a 32.1pp decline ($p$=0) in the probability of declaring ``Don't know'' when  $g_{jk,r}=0$ and a 46.6pp decline ($p$=0) when $g_{jk,r}=1$.  These large magnitudes suggests that $i$ has extremely limited knowledge when $j$ and $k$ are unreachable. Then, conditional on reachability, every extra step leads to a 10.9pp (or 14.8pp) ($p$=0) increase in the probability of declaring ``Don't know'' when there is no link (when there is a link) which again is a very large effect relative to the mean. 

Next, we further decompose the ``Don't know'' response into its two components --- not knowing that either $j$ or $k$ exist and knowing of $j$ and $k$, but not having a view on their relationship.  We explore this decomposition in Table \ref{tab:dontknow}.  Columns 1 and 2 consider whether respondent $i$ knows of both $j$ and $k$ as a function of distance and reachability.  Note that this variable is defined at the $(i,j,k)$ level and does not vary across relationship types. We therefore collapse the data in columns 1 and 2 to that level.  First note that $i$ knows of both $j$ and $k$ only 64\% of the time.  This means that approximately three-fourths of the ``Don't know'' responses come from not knowing anything about one or both of the nodes in question.  Unsurprisingly, the patterns of knowledge as a function of reachability and distance look very similar to those in Table \ref{tab:main}, Panel B.  In columns 3 and 4, we ask whether conditional on knowing that $j$ and $k$ exist, there is a residual relationship between $\text{DK}_{i,jk,r}$ and the distance between $i$ and $jk$.  We find that all of the qualitative patterns survive in this conditional regression: reachability is correlated with a lower likelihood of ``Don't know'', while larger social distances make a ``Don't know'' response more likely.  For example, when $g_{jk,r}=1$, moving $jk$ one step further from $i$ is associated with a 4.5pp increase ($p$=0) in $\text{DK}_{i,jk,r}$.\footnote{We note that the regressions in columns 3 and 4 control on an outcome that we have already shown to be correlated with distance.  See Section \ref{sec:conditional} below for a discussion of the likely bias.}

\subsubsection{Is there residual information conditional on a degree of certainty?}\label{sec:conditional}

Having established that the likelihood that a guess is correct and the likelihood of having a view both decline in distance, we now study whether these results are solely driven by respondents' uncertainty (``Don't know'') or whether within a certainty bin, $i$'s responses are more accurate when $jk$ are socially closer.  Table \ref{tab:main}, Panel C presents the results of regressions that condition on having a view about the relationship.  The control sets across columns are the same as those used in Panels A and B.

We first note that these regressions condition on individuals $i$, pairs $jk$, and relationship dimensions $r$ for which respondent $i$ claims to know about the relationship status.  Given that we just showed that having a view is also a function of distance, this creates a censoring problem that could lead to bias in the estimated relationship of interest. We believe that the most logical form of bias should make it harder to detect a negative relationship between social distance and bias.  This would be the case if $i$ were more likely to have an opinion about far $jk$ pairs who were nonetheless ``closer'' on unobservables and if those unobservables also corresponded to a higher likelihood of a correct guess.  

Keeping that caveat in mind, we do find evidence that distance is correlated with making a correct guess, even conditioning on $\text{DK}_{i,jk,r}=0$.  We again find that being 1 step higher in terms of average distance to $j$ and $k$ leads to a decrease of between 1.1 and 2.1pp ($p$=0.005) in the probability of guessing correctly when there is a link between $jk$. However, there is no detectable pattern in distance when $g_{jk,r}=0$. 

That network distance still predicts accuracy, even conditional on having a view, is consistent with the idea that simply knowing something about $j$ and $k$ does not drive the entire relationship identified in Table \ref{tab:main}, Panel A. Instead, a systematic bias develops more strongly outside the local radius.

\subsection{What best predicts network knowledge?}\label{sec:lasso}
Our findings thus far illustrate that network knowledge is systematically predicted by and related to the social distance between respondent $i$ and the pair $jk$ in question. We also demonstrated that this correlation is not affected by the inclusion of numerous controls including amenities, caste, geography, village fixed effects, and respondent fixed effects.

We now ask something stronger: out of all the variables we have at our disposal, what best predicts network knowledge?  For example, it may be geography, because people observe those who live nearby interacting with others, even if they are not themselves linked. Or it may be the network itself, because people interact on and through the network, so one is more likely to know about relationships among one's friends or friends of friends rather than someone who is an effective stranger.

To this end, we conduct a post-LASSO procedure (\citet{belloni2009least} and \citet*{belloni2014inference,belloni2014high}) by regressing the outcome, whether $i$ knows link $jk$ exists or not (or how certain $i$ is), on the network variables and all the demographic variables as before. Since this procedure penalizes putting in many parameters with the aim of selecting a sparse subset of covariates, it may wind up picking a minimal model that excludes network variables altogether. Alternatively, it may include network-based variables, viewing them as more predictive than any of our demographic covariates. Note that since LASSO is a shrinkage estimator, the post-LASSO procedure ensures that consistent estimates are being used.

Table \ref{tab:post-lasso} presents the results of this exercise. Column 1 contains the estimates when the outcome is whether the respondent's guess is correct ($y_{i,jk,r}$). Three core network variables are selected: the distance to the pair when there is no link, centrality of $i$ when $jk$ are linked, and the average centrality of the pair $jk$ when they are linked. As we have mentioned before, the centralities of $j$ and $k$ are highly (negatively) correlated with average distances with an arbitrary $i$.   Additionally (but omitted from the display), the following demographics are selected: dummies for whether $i$ and $j$, $i$ and $k$, and $j$ and $k$ are of the same caste, subcaste, have the same occupation, house ownership status, roof type, and electrification status.\footnote{23 village fixed effects and 1 dimension fixed effect are also selected.}  
In column 2 we predict the responses $\text{DK}_{i,jk,r}$ of the agent. We find that the distance is selected both when $g_{jk,r}=0$ and $g_{jk,r}=1$. Being one step further corresponds to a 8.99pp (or a 19.6\%) ($p$=0) increase in the probability of declaring ``Don't know''. Two other network measures are selected here: the centrality of the respondent when $g_{jk,r}=1$ and the average centralities of $jk$ when $g_{jk,r}=0$. In this case, a smaller set of demographics is selected: average geographic distance between $j$ and $k$, and all variables relating to the castes and subcastes of $i$, $j$, and $k$.\footnote{In addition, one occupation variable is selected along with 44 village fixed effects.} Note that in either case geographic distance from respondent to respondees or geographic centralities are not selected.

This suggests that an important set of predictive variables in terms of network knowledge or certainty about this knowledge consists of distance in the network from the respondent, how central the respondent is, and how central those in consideration are. These variables are selected by a post-LASSO procedure in a large horserace against a number of alternatives, including geographic variables. This is consistent with a story where people learn about their social structure by exploring/interacting through their social structure.

\

\section{Importance of assumptions on network knowledge}\label{sec:importance}

In this section, we discuss how our findings regarding limited network knowledge can be relevant for theoretical, applied, and econometric work. Assuming that agents have complete information about the underlying network is an oft-used assumption in the literature (with applications ranging from social learning and peer effects to network formation games). In what follows, we demonstrate how such an assumption (or lack thereof) can have first-order implications for a range of qualitative and quantitative results in various contexts.

\subsection{Social learning}\label{subsection:learning}

As our starting point, we explore the implications of network uncertainty in the context of learning over social networks. Many models of Bayesian social learning assume that agents have complete knowledge about the social network structure.\footnote{This includes both sequential learning models such as \citet*{banerjee1992simple}, \citet*{BHW}, and \citet*{SmithSorensen} as well as models of repeated network interactions such as \citet*{mossel2015strategic} and \citet*{MosselOlsmanTamuz}.} In what follows we use a simple framework to illustrate that relaxing the assumption of full network knowledge may lead to identification problems that can serve as impediments to learning, even when agents have access to enough information to uncover the state. Crucially, we also show that these identification problems do not arise when individuals face no uncertainty about the network. 

Consider a collection of $n+1$ individuals, denoted by $\{0, 1, \dots , n\}$, who wish to estimate an unknown state of the world $\theta\in\mathbb{R}$. Each agent $i$ receives a noisy private signal $s_i = \theta + \epsilon_i$ about the underlying state, where the error terms $\epsilon_i \sim \mathcal{N}(0,\sigma^2)$ are drawn independently across agents. Prior to observing their private signals, all agents share an (improper) uniform prior belief about the state. 

In addition to her private signals, each agent observes the point estimates of a subset of other individuals, whom we refer to as her neighbors. More specifically, we assume that agents are located on a directed, acyclic network that determines the patterns of observations: agent $i$ can observe agent $j$'s point estimate $\omega_j$ about the state if and only if there is a directed link from agent $j$ to agent $i$.\footnote{A directed network is said to be acyclic if it contains no directed cycles. The assumption that the network is acyclic ensures that information flows are unidirectional. This assumption is also equivalent to imposing an exogenous sequence of timing for agents' observations and assuming that each agent $i$ can only observe the point estimates of a subset of her predecessors, as in \citet*{banerjee1992simple} and \citet*{acemoglu2011bayesian}, among others.} 

We represent the potential lack of knowledge about the social network structure by assuming that agent $0$ is ex ante uncertain about the patterns of connections among other individuals. More specifically, we assume that the underlying network is drawn randomly from the set $G = \{\g^{1},\g^{2}\}$ with probabilities $p^1$ and $p^2 = 1-p^1$, respectively, where 
\begin{align*}
\g^i & =  \{ij: j \neq 1,2  \} \: \cup \:  \{j0:  j \neq 0\}
\end{align*}
and $jk$ denotes the directed edge from agent $j$ to agent $k$. Thus, whereas in both networks agent $0$ can observe the estimates of all other individuals, agents labeled $j = 3,\dots, n$ can only observe the estimate of agent $i\in \{1,2\}$ in network $\g^i$. The two networks are depicted in Figure \ref{figure:learning}. We have the following result.

\begin{proposition}\label{proposition:learning}
Suppose the social network is drawn randomly prior to the realization of the signals. As $n\to\infty$, 
\begin{enumerate}[(a)]
\item if agent $0$ observes the realized network, then she learns the state almost surely;
\item if agent $0$ has misspecified beliefs about the network, she mislearns the state almost surely;
\item if agent $0$ does not observe the realized network, she remains uncertain about the state almost surely.
\end{enumerate}
\end{proposition}

Statement (a) of the above result establishes that regardless of what the realized network and state are, agent $0$ uncovers the true state with arbitrarily high confidence as $n\to\infty$ as long as she has full knowledge about the underlying network. Note that even though $0$'s neighbors (partially) rely on a common source of information, agent $0$ can use her knowledge about the network structure to account for any redundancies in her neighbors' estimates. More specifically, when the realized network is $\g^i$, the point estimate of each agent $j \in\{ 3,\dots, n\}$ is equal to $\omega_j = (s_j+s_i)/2$, while $\omega_1 = s_1$ and $\omega_2 = s_2$. Hence, with access to the point estimates of agents 1 and 2, agent $0$ can simply back out agent $j$'s private signal $s_j$ by computing $2\omega_j - \omega_i$. Such a calculation enables agent $0$ to discern any correlation in her neighbors' estimates that is due to their common neighbor $i$. Consequently, agent $0$'s point estimate of the true state is given by
\begin{align*}
\omega_0 & = \frac{1}{n+1}\left(s_0 + \omega_1 + \omega_2 + \sum_{j=3}^n (2\omega_j-\omega_i)\right) = \frac{1}{n+1}\sum_{j=0}^{n} s_j, 
\end{align*}
which by the law of large numbers, converges to the true state with probability one as $n\to\infty$.

Contrasting the above observation with statements (b) and (c) of Proposition \ref{proposition:learning} underscores the crucial role of \textit{correct} and \textit{complete} network knowledge for successful learning in part (a): whereas agent $0$ learns the state with probability one when she observes the realized network, she fails to learn the state if she has either incorrect or incomplete knowledge about the realized network. 

To see the intuition for this contrast, first consider the case in which agent $0$ has misspecified beliefs about the social network (statement (b)). More specifically, suppose without loss of generality that the true realized network is $\g^1$, but agent $0$ believes (mistakenly) that the underlying network is $\g^2$. Such a misspecification means that even though the point estimate of agent $j \in \{3,\dots, n\}$ is given by $\omega_j = (s_j+s_1)/2$, agent $0$ assumes that the correlation in her neighbors' estimates are due to agent 2's private signal. As a result, agent $0$'s point estimate is given by 
\begin{align*}
 \omega_0 & = \frac{1}{n+1}\left(s_0 + \omega_1 + \omega_2 + \sum_{j=3}^n (2\omega_j-\omega_2)\right) = \frac{1}{n+1} \left((n-2)(s_1-s_2) + \sum_{j=0}^n s_j \right),
\end{align*}
which converges to $\theta + s_1 - s_2$ with probability one, whenever the true state is $\theta$. In other words, agent $0$'s misspecified belief about the network structure manifests itself as a bias in her estimate about the state. 

Finally, part (c) of Proposition \ref{proposition:learning} illustrates that uncertainty about the network structure can create an identification problem that would serve as an impediment to learning. In the context of the above example, no matter how many neighbors agent $0$ has, she can only learn the state by properly accounting for the redundancies and patterns of correlations in her neighbors' estimates. Yet, uncertainty about the identity of her neighbors' neighbors means that agent $0$ cannot identify the source of such redundancies. More specifically, when the true state and the realized network are $\theta$ and $\g^{1}$, respectively, agent $0$ has no way of distinguishing whether the underlying state-network pair is $(\theta,\g^1)$ or $(\theta + s_1-s_2, \g^2)$, even as $n\to\infty$. This is despite the fact that she can hold and update beliefs not only about the state but also about the underlying network structure.\footnote{In this sense, our result on the role of network uncertainty is similar to \citet*{DaronFragility}, who argue that uncertainty about the signal generating process can result in an identification problem for a single Bayesian agent in isolation, as the same long-run frequency of signals may be consistent with multiple states.}

Taken together, Proposition \ref{proposition:learning} illustrates that incorrect or incomplete knowledge about the social network can undermine information aggregation, even in fairly simple environments that would have otherwise led to learning. This observation, alongside our empirical findings in Section \ref{sec:network_distribution}, suggests that Bayesian models that allow for uncertainty about the network structure (such as \citet*{LobelSadlerTE}) as well as behavioral models (such as \citet*{degroot1974reaching}, \citet*{EysterRabinQJE}, and  \citet*{li2017learning})  that assume agents cannot fully account for how the network structure shapes their observations may be useful and important starting points for applied work to focus on.


\subsection{Network interactions and peer effects}\label{subsection:BCZ}

We next focus on network peer effect games and illustrate how assuming full network knowledge may lead to biased structural estimates when in fact agents face uncertainty about the underlying network. 

To this end, we focus on the canonical network interaction model of  \citet*{ballester2006s}. Besides their widespread use, models that build off this linear-quadratic framework lend themselves to readily obtainable structural estimates, as well as counterfactual and policy analyses (such as which ``key'' players to target with an intervention). Applications include peer effects in education \citep*{CalvoREStud}, local consumption externalities \citep*{Ozan}, research collaboration among firms \citep*{GoyalRAND}, among numerous others.

The basic setup is an $n$-player game in which agents' payoffs are given by 
\begin{align*}
u_i(x_1,\dots, x_n) & =  x_i\theta_i -\frac{1}{2}x_i^2 +  \alpha \sum_{j\neq i} g_{ij}x_ix_j,
\end{align*}
where $x_i$ denotes the action (say, effort) of agent $i$, $g_{ij}$ regulates the strength of strategic complementarities between $i$ and $j$, and parameters $\alpha \in [0,1)$ and $\theta_i>0$ are common knowledge among all agents. 

As is common in the literature, pairwise interactions can be represented by a directed and weighted network $\g = [g_{ij}]$, with the convention that $g_{ii}=0$ for all $i$. In a departure from the rest of the literature, however, we assume that this interaction network is drawn randomly prior to the start of the game. More specifically, we assume that for any pair of agents $i\neq j$, the weight $g_{ij}$ is drawn independently from the rest of the weights according to a non-degenerate probability distribution with density $f_{ij}$. Throughout, we assume that $\sum_{j\neq i} \max \supp(f_{ij}) \leq 1$ for all agents $i$.\footnote{This restriction on the support of the distributions is meant to guarantee that the matrix $I-\alpha \g$ is invertible with probability one.} 

We consider two variants of the above game. In what we call the complete information benchmark, we assume that the realized network $\g$ is observed by all agents prior to making their decisions. Therefore, ex post, our complete information benchmark coincides with the canonical model of \citet*{ballester2006s}. On the other hand, under the incomplete information variant of the game, we assume that each agent $i$ can only observe the set of weights $\{g_{ij}: j \neq i\}$ in her neighborhood. Hence, in contrast to the complete information benchmark, agents choose their actions with no knowledge about the strength of strategic complementarities among other agents. We have the following result:

\begin{proposition}\label{proposition:BCZ}
The vectors of equilibrium actions under the complete and incomplete information variants of the game are given by 
\begin{align}
x^{\mathrm{com}} & = (I-\alpha \g)^{-1}\theta \label{eq:complete}\\
x^{\mathrm{inc}} & = \left(I+\alpha \g (I-\alpha \mathbb{E}[\g])^{-1}\right)\theta, \label{eq:incomplete}
\end{align}
respectively. Furthermore, $\mathbb{E}[x_i^{\mathrm{inc}}] < \mathbb{E}[x_i^{\mathrm{com}}]$ for all agents $i$. 
\end{proposition}

The above result establishes that equilibrium actions are sensitive to whether agents have full knowledge about the patterns of interactions in the network. More importantly however, Proposition \ref{proposition:BCZ} also illustrates that uncertainty about the network structure results in a systematic shift in equilibrium actions: each agent's action in the incomplete information game is in expectation less than her  action in the complete information benchmark. 

What this means, practically, is that estimates of $\theta$ and $\alpha$ obtained under the assumption of complete information may be biased if agents are in fact uncertain about the network structure. Such biases can in turn substantially affect not only counterfactuals but also policy prescriptions by potentially impacting (i) cost-benefit assessments, (ii) the predicted extent of externalities, and (iii) the relative network centralities of various agents; a statistic that is central to policy prescriptions in this literature. 


\subsection{Identification in network formation games}\label{subsection:formation}

We conclude this section by investigating the implications of incomplete network knowledge for identification of structural parameters in network formation games. As our main result, we show that introducing the (arguably more realistic) assumption of incomplete information may transform a model that is fundamentally hard to identify to a straightforwardly estimable model. 

Consider a simultaneous-move game of network formation with $n$ players, indexed $\{1,\dots, n\}$. Each agent $i$ is endowed with some attribute $z_i \in \{a,b\}$. These attributes, which are common knowledge among all agents and are observable to the econometrician, may represent agents' gender, caste, race, or other characteristics. Throughout, we use $\lambda$ to denote the fraction of agents with attribute $a$. 

Agents draw utilities by forming direct and indirect links with other individuals. More specifically, the utility function of agent $i$ is given by 
\begin{align}\label{eq:payoffs}
u_i(\g, z, \epsilon_{i}) & =  \sum_{j\neq i} g_{ij}(\alpha-\epsilon_{ij}) + \beta \sum_{j\neq i} g_{ij} \mathbf{1}\{z_j = z_i\} + \gamma \left(\frac{1}{n-1 }\sum_{j\neq i}\sum_{k\neq i,j} g_{ij}g_{jk}\right),
\end{align}
where $\g$ denotes the undirected, unweighted network of linkages (with $g_{ij} = 1$ if $i$ and $j$ are linked and $g_{ij} = 0$ otherwise), $z = (z_1,\dots,z_n)$ denotes the vector of population attributes, and $\epsilon_{ij}$ is a preference shock that determines agent $i$'s idiosyncratic cost of establishing a link to agent $j$. Throughout, we assume that preference shocks are drawn independently from a common probability distribution $F(\cdot)$ with a continuous density and full support over $\mathbb{R}$. 

The first two terms on the right-hand side of \eqref{eq:payoffs} capture the (net) \textit{direct} utility of forming linkages: agent $i$ obtains a net marginal utility of $\alpha - \epsilon_{ij}$ of forming a direct link to agent $j$, with an additional marginal benefit of $\beta$ if the two agents have identical attributes. On the other hand, the third term on the right-hand side of \eqref{eq:payoffs} allows for the possibility that \textit{indirect} connections can also be payoff relevant. For example, when $\gamma> 0$, agent $i$ draws a benefit if any of her neighbors form links with other individuals.\footnote{As in \citet*{sheng2016structural} and \citet*{Mele2017}, the specification in \eqref{eq:payoffs} assumes that the indirect benefits of forming links are proportional to the number of connections of $i$'s neighbors. An alternative specification is to assume that indirect benefits only depend on the number of \textit{distinct} connections of $i$'s neighbors, as in \citet*{de2017identifying}. Our choice of the specification in \eqref{eq:payoffs} is made to simplify the analysis. However, our main result on the relationship between network knowledge and the identifiability of structural parameters holds for a broader class of specifications.} 

Players simultaneously announce the set of agents they wish to link to, with links formed by mutual consent. Formally, each player $i$ chooses the action $x_{ij}\in \{0,1\}$ indicating whether she wishes to form a link with agent $j$. The undirected link $(i,j)$ is then formed if both parties agree, i.e., $g_{ij} = x_{ij}x_{ji}$. 

The econometrician is interested in estimating the structural parameters $\alpha$, $\beta$, and $\gamma$ in \eqref{eq:payoffs} while only observing the equilibrium network $\g$.  The literature typically assumes that the realizations of preference shocks $\epsilon_{ij}$ are common knowledge among the agents and that agents are in a pairwise stable network  \citep*{de2017identifying,leung2014random,menzel2015strategic,sheng2016structural}. This observability assumption means that agents' link formation decisions $x_{ij} \in \{0,1\}$ can be contingent on the realization of all pairwise preference shocks.

The fact that network formation games of complete information tend to pose a challenge for identification has been at the forefront of the literature. To overcome this issue, the literature has resorted to imposing subsequent stronger assumptions on agents' payoffs, the number of preference shocks, or the network formation process (see, e.g.,  \citet{de2017identifying,leung2014random,menzel2015strategic,sheng2016structural}). 

In a departure from the literature, we assume that preference shocks are only locally observable: the realization of agent $i$'s preference shocks $\epsilon_i = (\epsilon_{i1}, \dots, \epsilon_{in})$ are $i$'s private information and hence are unobservable to all other agents. 
 Given the incomplete information nature of this specification, we focus on (symmetric monotone) Bayes-Nash equilibrium as our solution concept, according to which each agent $i$ chooses the action $x_{ij} =1$ if the realized preference shock $\epsilon_{ij}$ falls below some given threshold that only depends on $i$ and $j$'s attributes.

Before presenting our result, we remark that the notion of incomplete information in this context is slightly different from those in Subsections \ref{subsection:learning} and \ref{subsection:BCZ}. Whereas our earlier examples focused on the observability (or lack thereof) of the network structure itself, the incomplete information variant of the network formation game above requires the agents to be uncertain about the realizations of the underlying pairwise preference shocks.

\begin{proposition}\label{proposition:formation}
Suppose the econometrician observes the realized network but not the realization of the preference shocks  and  $\lambda\neq 1/2$. If preference shocks are only locally observable, structural parameters $\alpha$, $\beta$, and $\gamma$ are point-identified. 
\end{proposition}

 The proposition illustrates that relaxing the complete information assumption leads to a diametrically opposite result
 of that in the literature: observing the pattern of linkages provides the analyst with sufficiently rich information to point-identify all structural parameters. This contrast is driven by the fact that when players have limited information, they are unable to condition their decisions on the fine details of the entire set of realized shocks $\{\epsilon_{ij}\}$, leading to coarser linking strategies and hence facilitating identification. This illustrates that imposing the simple and --- in view of our empirical findings in Section \ref{sec:network_distribution} --- more realistic assumption of incomplete information transforms a model that is fundamentally hard to identify to a realistic and straightforwardly estimable model, without changing the payoffs, imposing specific link formation dynamics, or restricting the number or types of links that can be formed.

\section{Conclusion}\label{sec:conclusion}

We present a new set of stylized facts about the extent to which individuals are informed about network relationships in their own communities.  We collected survey data in 75 communities where we had already documented the full network structure across a number of dimensions.  We find that network knowledge is low and localized: respondents often do not know linking statuses among others in their village and that network knowledge decays steeply in network distance to the pair. These patterns are robust to the inclusion of numerous covariates (e.g., amenities, caste, geography) and are consistent with individuals learning about the social network through the network itself.

Our findings are at odds with many commonly-used modeling techniques in network economics across a range of questions and applications. Moving from a complete information framework to an incomplete information framework can lead to changes in the behavior of agents in the models and, thus different predictions or equilibria, thereby affecting qualitative conclusions and structural estimates.  Further, we show that typical econometric models of network formation have the features that moving to the more realistic assumption of incomplete information may make parameters identifiable whereas under the less realistic assumption of complete information, parameters may not be identifiable or only partially identifiable. We are certainly not the first to explore incomplete information in any of these cases. However, we view our examples playing a pedagogical role, pointing out how this ubiquitous assumption in the literature, inconsistent with the data, may affect widespread conclusions.

Taken together, our empirical and theoretical results suggest that the complete information benchmark is not an accurate view of the world for the setting we study, and moreover, is not always a benign assumption.  Rather, incorporating incomplete network information may yield theoretical predictions that better match the behaviors of agents  in real-world networks. Incomplete information may also play a simplifying role in the theory, both yielding a smaller set of equilibria in games of cooperation and also leading to smaller identified sets in network econometric models.

\newpage
\bibliographystyle{ecta}
\bibliography{networks}

\newpage

\appendix

\input{Figures_Tables_paper}

\newpage

\clearpage

\section{Proofs}\label{sec:proof}
\input{ProofAppendix}

\clearpage

\begin{center}
\large{{\bf ONLINE APPENDIX\\ NOT FOR PUBLICATION}}
\end{center}

\section{Main Tables with Centralities Displayed}\label{sec:w_cent}
\setcounter{table}{0}
\renewcommand{\thetable}{B.\arabic{table}}
\setcounter{figure}{0}
\renewcommand{\thefigure}{B.\arabic{figure}}

\input{apdx_w_cent}

\end{document}

%% file: Figures_Tables_paper.tex
\section*{Figures}

{\centering
\begin{figure}[!h]
\vspace*{-0.75cm}
\hspace*{-1cm}
\subfloat[Share of Correct Guesses by Distance]{
\includegraphics[scale = 0.4]{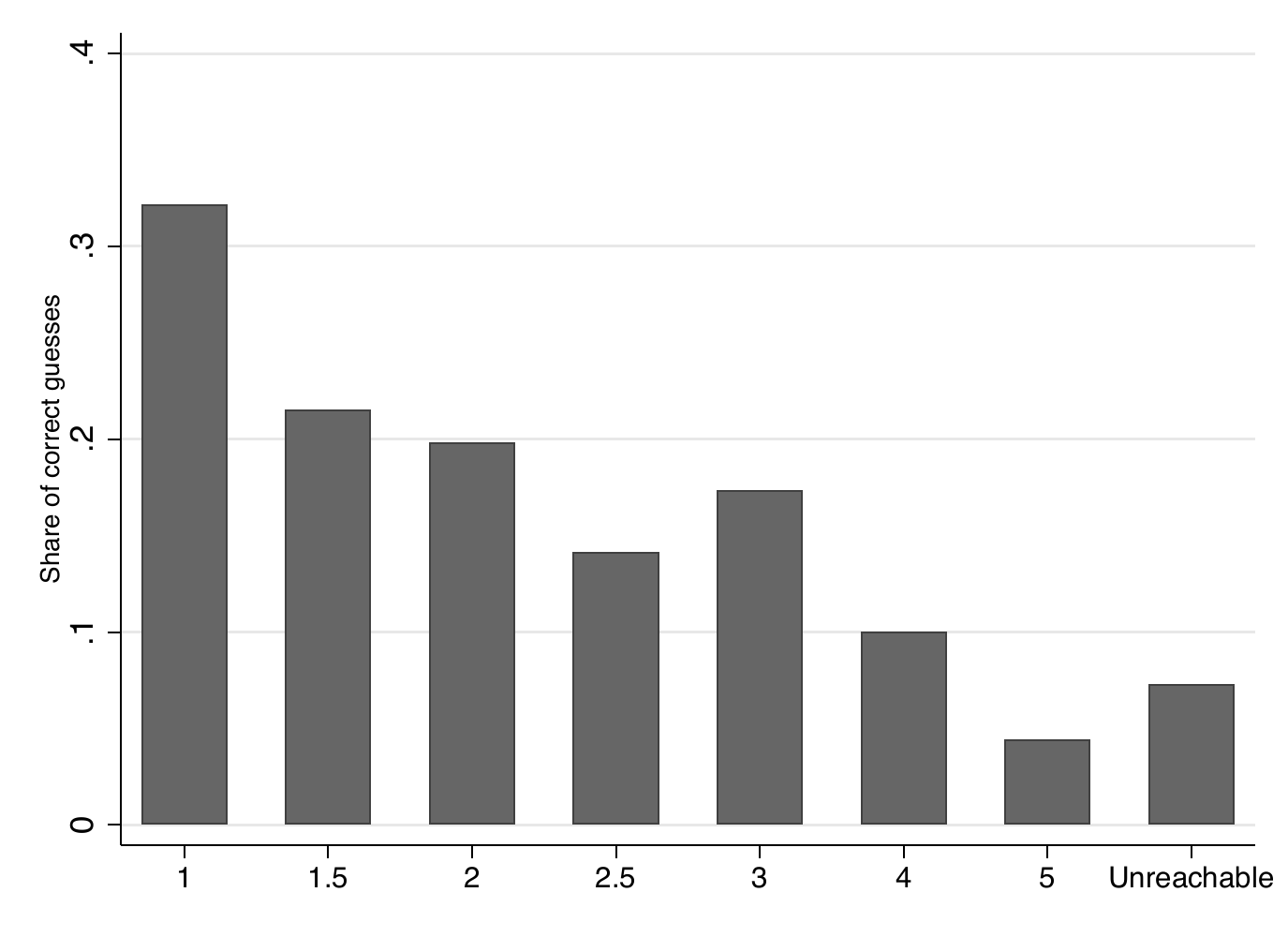}
}
\subfloat[Share of Don't Knows by Distance]{
\includegraphics[scale = 0.4]{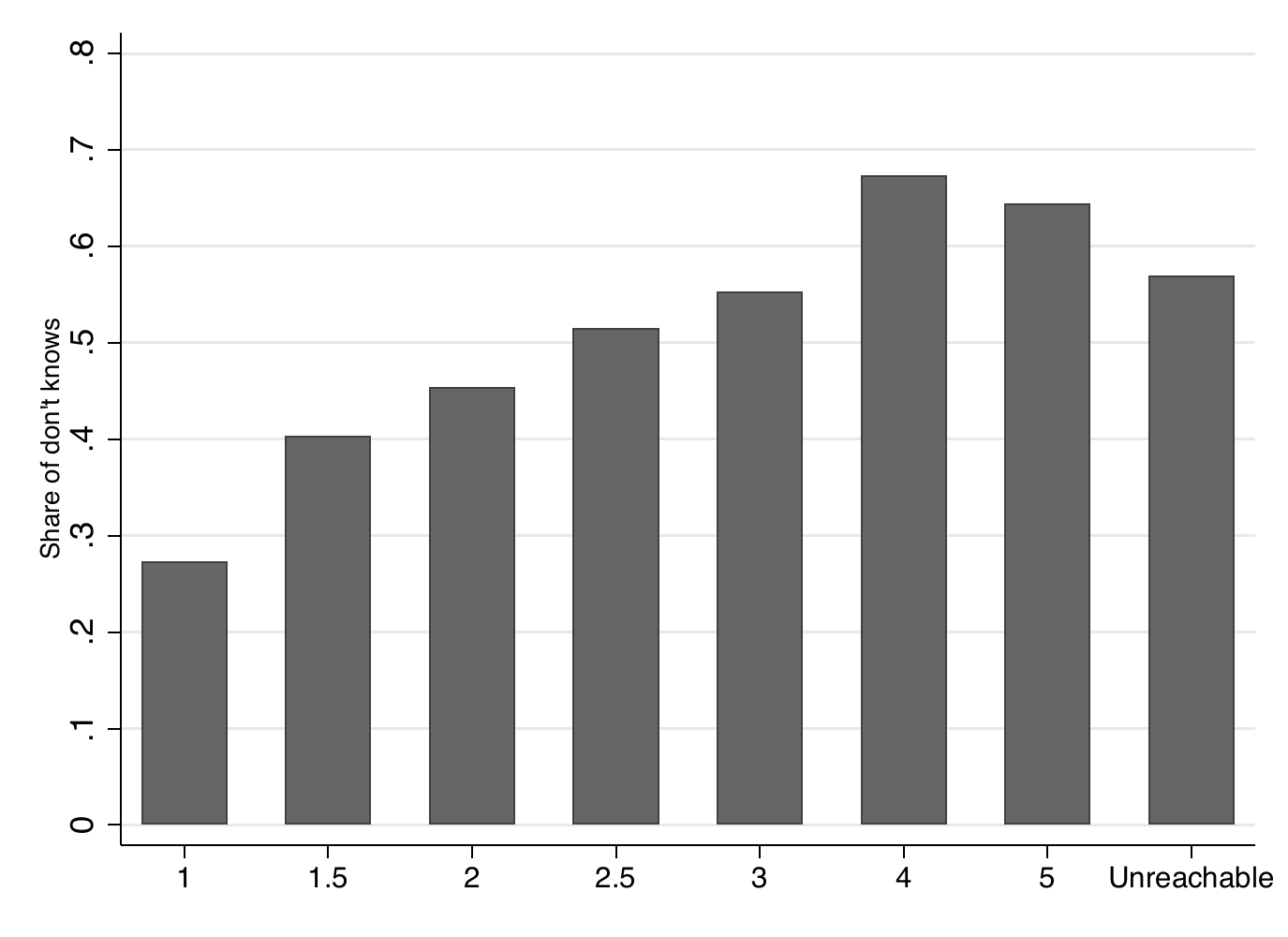}
}
\subfloat[Share of Correct Guesses by Distance, conditional on claiming to know]{
\includegraphics[scale = 0.4]{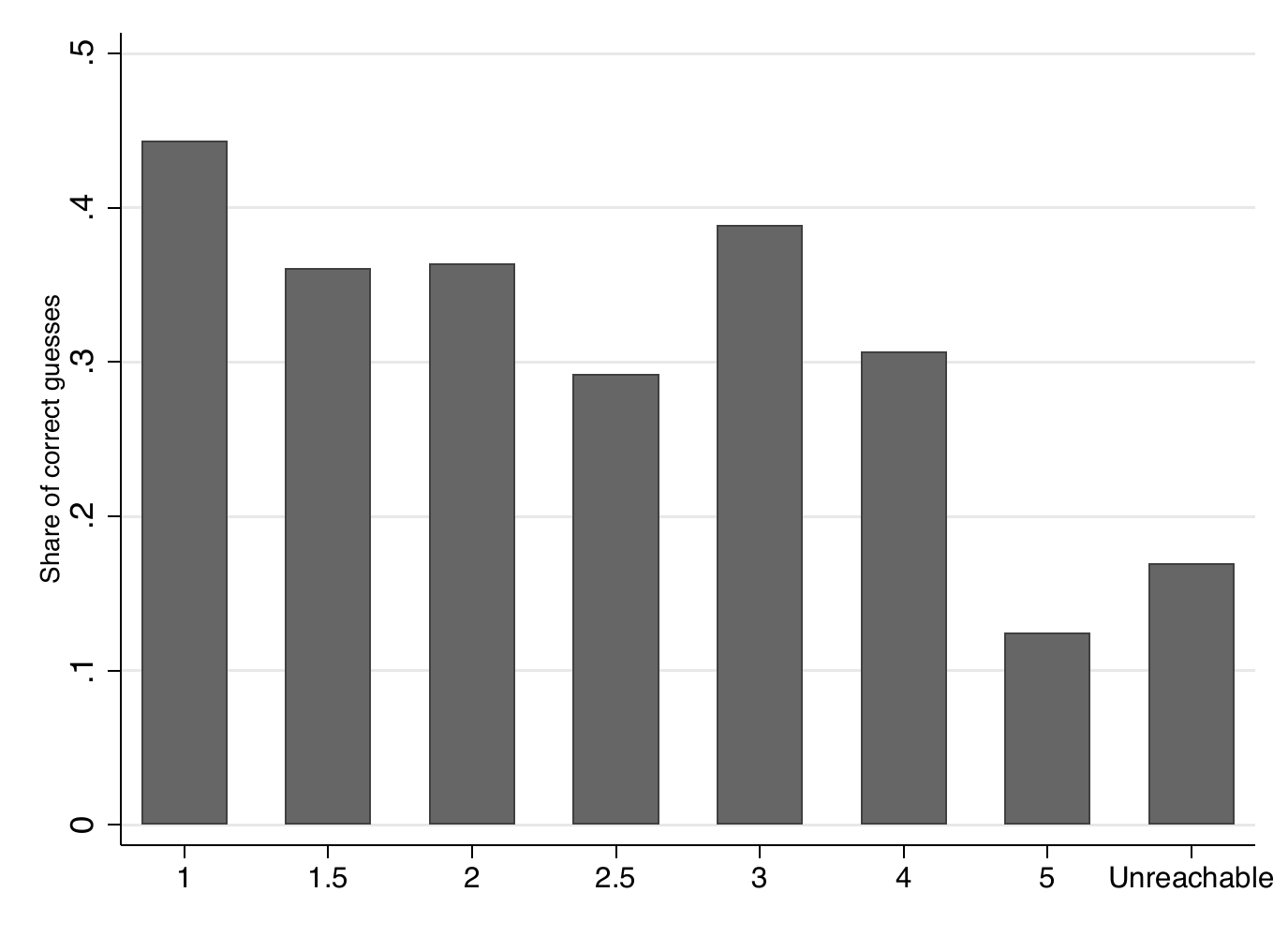}
}

\hspace*{-1cm}
\subfloat[Share of Correct Guesses by Centrality of $jk$]{
\includegraphics[scale = 0.4]{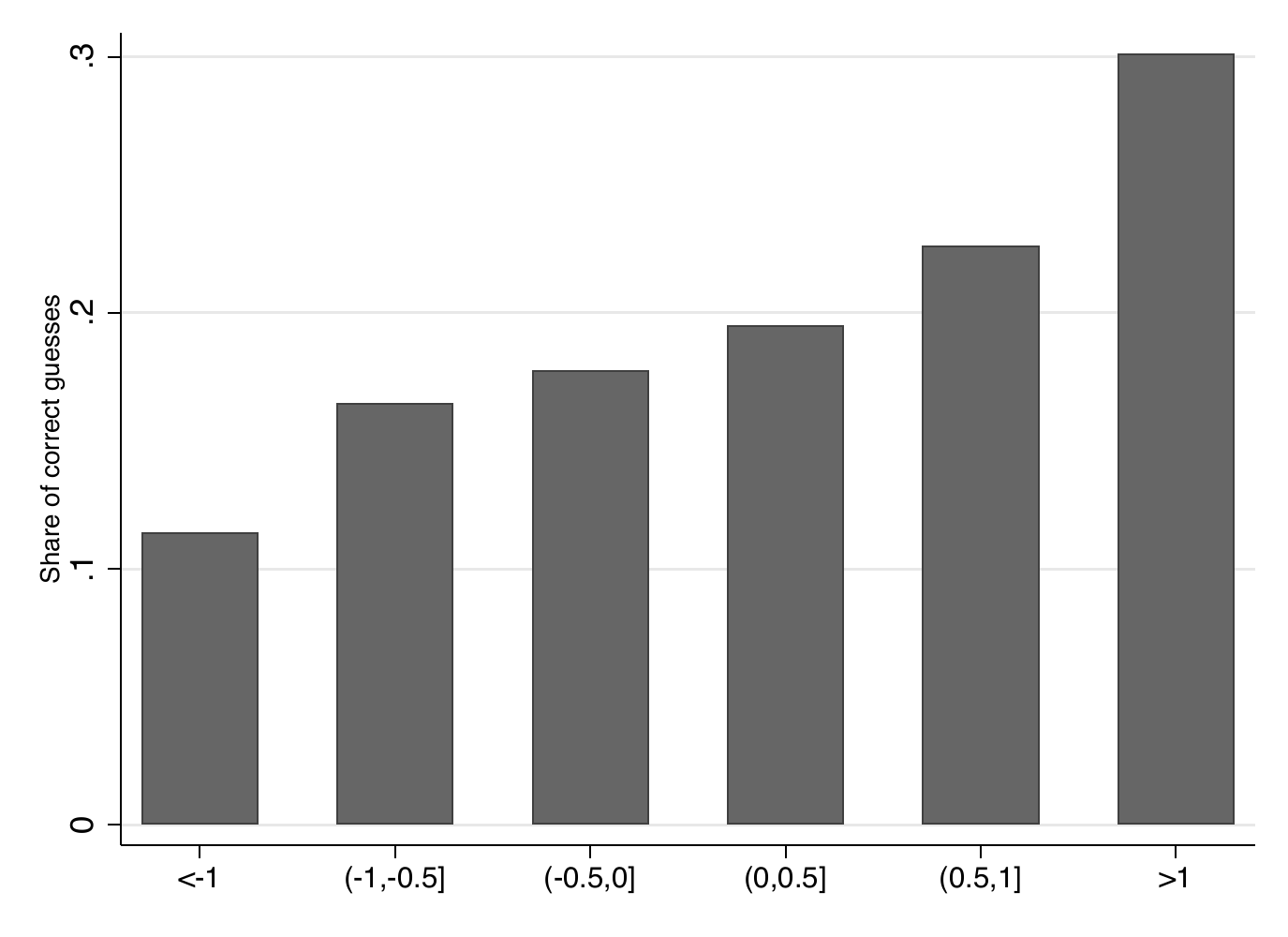}
}
\subfloat[Share of Don't Knows by Centrality of $jk$]{
\includegraphics[scale = 0.4]{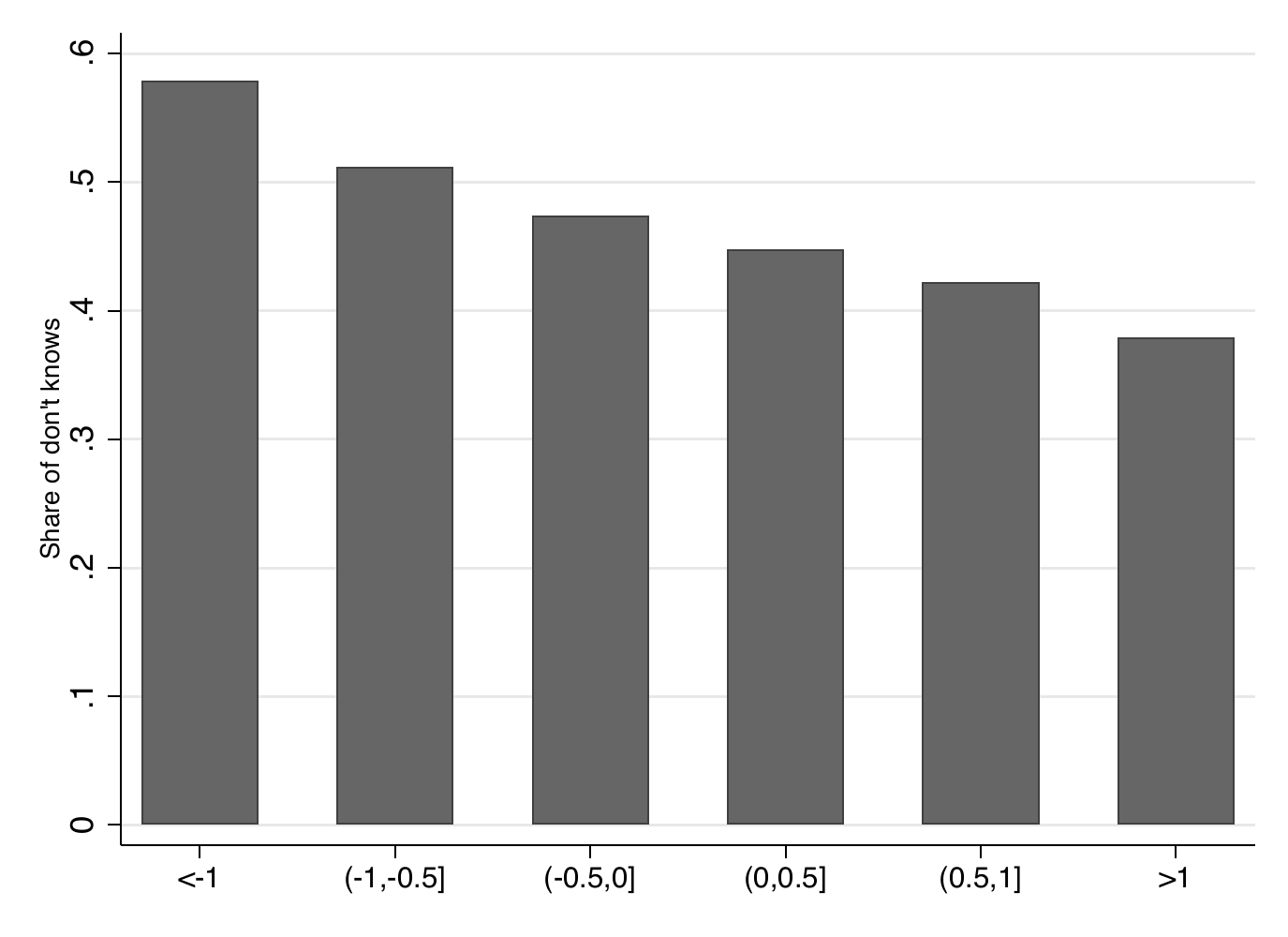}
}
\subfloat[Share of Correct Guesses by Centrality of $jk$, conditional on claiming to know]{
\includegraphics[scale = 0.4]{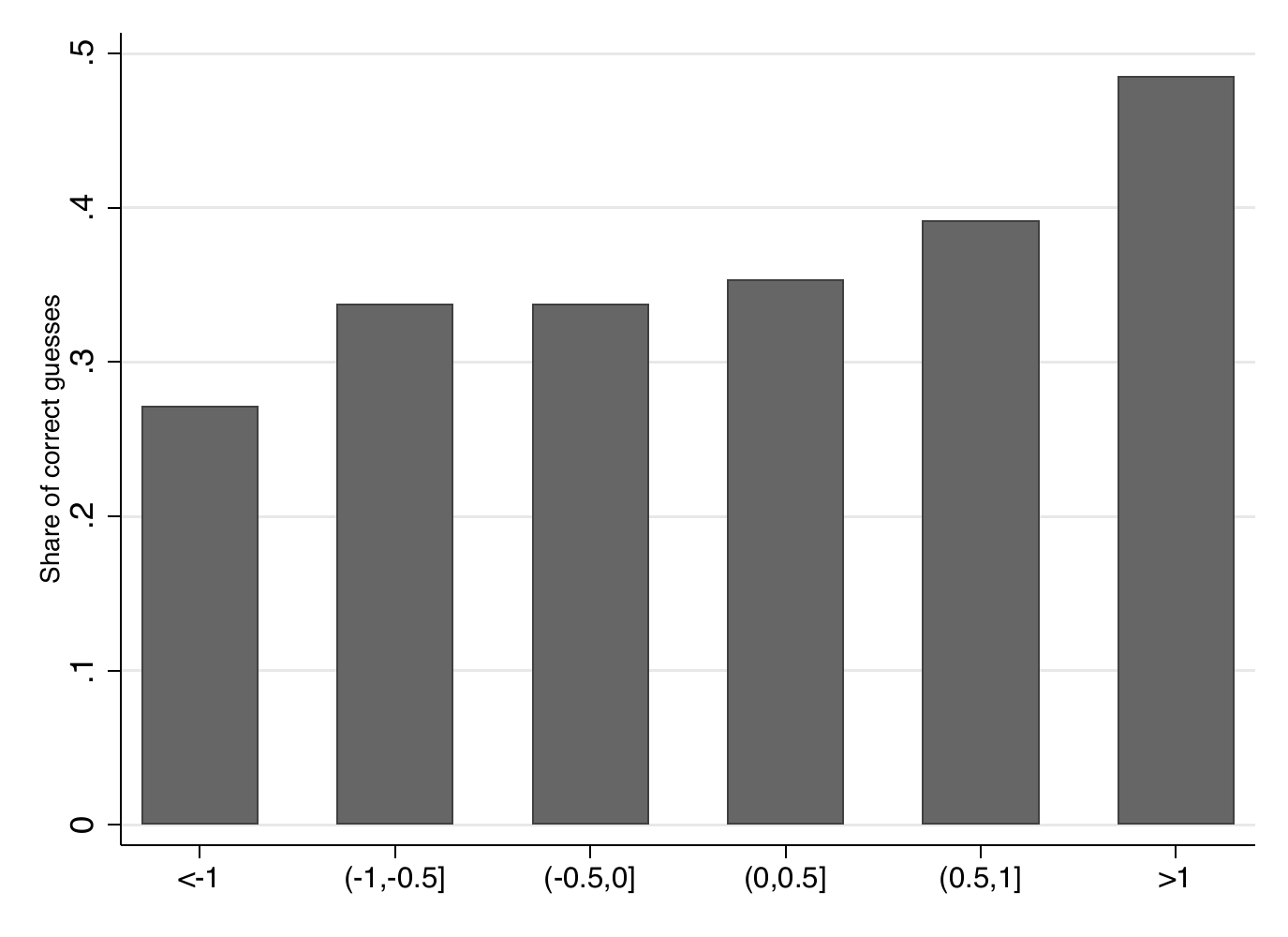}
}

\caption{Panels A--C depict raw shares of correct guesses and declaring that the respondent doesn't know by distance. 
 Panels D--F show the same by average centrality of $j$ and $k$ (standardized). 
 \label{fig:raw}}

\hfill
\end{figure}
}

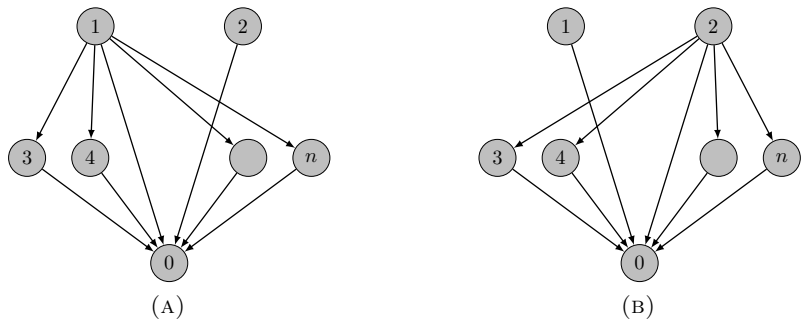
\begin{figure}[h!]
\subfloat[]{
\scalebox{0.7}{\begin{tikzpicture}
	\def \n {5}
	\def \radius {1.7cm}
	\def \margin {8}
	\node[draw, circle, fill=gray!50, minimum size=20pt] at (0,-2) (v0){$0$};
	\node[draw, circle, fill=gray!50, minimum size=20pt] at (2.7,0) (vn){$n$};
	\node[draw, circle, fill=gray!50, minimum size=20pt] at (1.5,0) (v5){};
	\node[draw, circle, fill=gray!50, minimum size=20pt] at (-1.5,0) (v4){$4$};	
	\node[draw, circle, fill=gray!50, minimum size=20pt] at (-2.7,0) (v3){$3$};
	\node[draw, circle, fill=gray!50, minimum size=20pt] at (1.4,2.5) (v2){$2$};
	\node[draw, circle, fill=gray!50, minimum size=20pt] at (-1.4,2.5) (v1){$1$};
	\draw[line width = 0.25mm, ->, >=latex] (v1) to (v3);
	\draw[line width = 0.25mm, ->, >=latex] (v1) to (v4);
	\draw[line width = 0.25mm, ->, >=latex] (v1) to (v5);
	\draw[line width = 0.25mm, ->, >=latex] (v1) to (vn);
	\draw[line width = 0.25mm, ->, >=latex] (v2) to (v0);
	\draw[line width = 0.25mm, ->, >=latex] (v3) to (v0);
	\draw[line width = 0.25mm, ->, >=latex] (v4) to (v0);
	\draw[line width = 0.25mm, ->, >=latex] (v5) to (v0);
	\draw[line width = 0.25mm, ->, >=latex] (vn) to (v0);
	\draw[line width = 0.25mm, ->, >=latex] (v1) to (v0);	
\end{tikzpicture}
}}\hspace{0.64in}
\subfloat[]{
\scalebox{0.7}{\begin{tikzpicture}
	\def \n {5}
	\def \radius {1.7cm}
	\def \margin {8}
	\node[draw, circle, fill=gray!50, minimum size=20pt] at (0,-2) (v0){$0$};
	\node[draw, circle, fill=gray!50, minimum size=20pt] at (2.7,0) (vn){$n$};
	\node[draw, circle, fill=gray!50, minimum size=20pt] at (1.5,0) (v5){};
	\node[draw, circle, fill=gray!50, minimum size=20pt] at (-1.5,0) (v4){$4$};	
	\node[draw, circle, fill=gray!50, minimum size=20pt] at (-2.7,0) (v3){$3$};
	\node[draw, circle, fill=gray!50, minimum size=20pt] at (1.4,2.5) (v2){$2$};
	\node[draw, circle, fill=gray!50, minimum size=20pt] at (-1.4,2.5) (v1){$1$};
	\draw[line width = 0.25mm, ->, >=latex] (v2) to (v3);
	\draw[line width = 0.25mm, ->, >=latex] (v2) to (v4);
	\draw[line width = 0.25mm, ->, >=latex] (v2) to (v5);
	\draw[line width = 0.25mm, ->, >=latex] (v2) to (vn);
	\draw[line width = 0.25mm, ->, >=latex] (v2) to (v0);
	\draw[line width = 0.25mm, ->, >=latex] (v3) to (v0);
	\draw[line width = 0.25mm, ->, >=latex] (v4) to (v0);
	\draw[line width = 0.25mm, ->, >=latex] (v5) to (v0);
	\draw[line width = 0.25mm, ->, >=latex] (vn) to (v0);
	\draw[line width = 0.25mm, ->, >=latex] (v1) to (v0);	
\end{tikzpicture}
}}
\caption{\small Networks $\g^1$ (left) and $\g^2$ (right). Arrows indicate the direction over which information can flow from one agent to another.\label{figure:learning}}
\label{figure:regular}
\end{figure}

\clearpage

\section*{Tables}

\begin{table}[!h]
\centering
\caption{Sample Statistics \label{tab:summary}}
\scalebox{1}{\begin{threeparttable}
\emph{Panel A: Respondent Characteristics}
\input{table_respondent_summary_ed.tex}
\vspace{0.05in}
\emph{Panel B: Pair Characteristics}
\input{table_pair_summary.tex}
\vspace{0.05in}
\emph{Panel C: Knowledge}
\input{table_knowledge_summary_ed.tex}
\begin{tablenotes}
\end{tablenotes}
\end{threeparttable}}
\end{table}

\clearpage

\begin{table}[!h]
\centering
\caption{Network Knowledge as a function of Distance and Centralities \label{tab:main}}
\scalebox{0.78}{\begin{threeparttable}
\emph{Panel A: Correct guesses (including don't knows as wrong guesses)}
\input{table_reachable_panel_dim5.tex}
\emph{Panel B: Don't knows}
\input{table_reachable_dontknow_panel_dim5.tex}
\emph{Panel C: Correct guesses (conditional on sure guesses only)}
\input{table_reachable_know_panel_dim5.tex}
\begin{tablenotes}
Notes: Standard errors (clustered at village level) are reported in parentheses. $p$-values are reported in brackets. All columns control for eigenvector centrality of $i$ and avg. eigenvector centrality of $j$,$k$, when $g_{jk}$ is 1 and 0. All columns also include dimension fixed-effects. Columns (2), (3) and (4) control for demographic covariates. Demographic controls include average geographic distance of $i$ from $j$ and $k$ and dummies for each of whether $i$ is of the same caste, subcaste, has the same electrification status, has the same roof type, has the same number of rooms, and has the same house ownership status for each of $j$ and $k$, when $g_{jk}$ is 1 and 0. They also include the above-mentioned dummies for when $j$ and $k$ have the same demographic traits. Columns (3) and (4) include village and respondent fixed-effects, respectively.
\end{tablenotes}
\end{threeparttable}}
\end{table}

\clearpage

\begin{table}[!h]
\centering
\caption{Network Knowledge, conditional on knowing $j$,$k$, as a function of Distance and Centralities \label{tab:dontknow}}
\scalebox{1}{\begin{threeparttable}
\input{table_knowpair_panel_dim5.tex}
\begin{tablenotes}
Notes: Standard errors (clustered at village level) are reported in parentheses. $p$-values are reported in brackets. All columns control for eigenvector centrality of $i$ and avg. eigenvector centrality of $j$,$k$, when $g_{jk}$ is 1 and 0. Columns (3) and (4) include dimension fixed-effects. Columns (2) and (4) control for demographic covariates. Demographic controls include average geographic distance of $i$ from $j$ and $k$ and dummies for each of whether $i$ is of the same caste, subcaste, has the same electrification status, has the same roof type, has the same number of rooms, and has the same house ownership status for each of $j$ and $k$, when $g_{jk}$ is 1 and 0. They also include the above-mentioned dummies for when $j$ and $k$ have the same demographic traits.
\end{tablenotes}
\end{threeparttable}}
\end{table}


\begin{table}[!h]
\centering
\caption{Post-LASSO Estimates \label{tab:post-lasso}}
\scalebox{1}{\begin{threeparttable}
\input{table_lasso_panel_dim5.tex}
\begin{tablenotes}
Notes: Standard errors (clustered at village level) are reported in parentheses. $p$-values are reported in brackets. Demographic controls include average geographic distance of $i$ from $j$ and $k$ and dummies for each of whether $i$ is of the same caste, subcaste, has the same electrification status, has the same roof type, has the same number of rooms, and has the same house ownership status for each of $j$ and $k$, when $g_{jk}$ is 1 and 0. They also include the above-mentioned dummies for when $j$ and $k$ have the same demographic traits. Post-LASSO procedure used with optimal penalty parameter to select variables.
\end{tablenotes}
\end{threeparttable}}
\end{table}

\clearpage

%% file: table_respondent_summary_ed.tex
{
\def\sym#1{\ifmmode^{#1}\else\(^{#1}\)\fi}
\begin{tabular*}{10cm}{@{\hskip\tabcolsep\extracolsep\fill}l*{1}{ccc}}
\hline\hline
                                   &     &        mean&         obs\\
\hline
Female                             &    &        0.59&        4838\\
Age                                &    &       42.55&        4838\\
General or OBC                     &     &        0.65&        4838\\
Born in the village                    &     &        0.27&        4838\\
Born in the village (female)           &     &        0.11&        2854\\
Born in the village (male)             &     &        0.51&        1984\\
Degree                  &     &       19.51&        4838\\
Eigenvector centrality             &     &        0.07&        4838\\
\hline\hline
\end{tabular*}
}

%% file: table_pair_summary.tex
{
\def\sym#1{\ifmmode^{#1}\else\(^{#1}\)\fi}
\begin{tabular*}{10cm}{@{\hskip\tabcolsep\extracolsep\fill}l*{1}{cc}}
\hline\hline
                                        &        mean&         obs\\
\hline
j,k Reachable                           &        0.99&       21861\\
j,k Linked                              &        0.34&       21861\\
Avg. distance to j,k                    &        2.13&       21861\\
Avg. degree centrality of j,k           &       18.68&       21861\\
Avg. eigenvector centrality of j,k      &        0.06&       21861\\
Same caste category j,k                 &        0.59&       21834\\
Same subcaste j,k                       &        0.44&       21818\\
\hline\hline
\end{tabular*}
}

%% file: table_knowledge_summary_ed.tex
{
\def\sym#1{\ifmmode^{#1}\else\(^{#1}\)\fi}
\begin{tabular*}{10cm}{@{\hskip\tabcolsep\extracolsep\fill}l*{1}{ccc}}
\hline\hline
                                     &   &        mean&         obs\\
\hline
Correct guess                         &  &        0.20&       66810\\
Correct guess ($j$,$k$ not linked)        &  &        0.07&       43884\\
Correct guess ($j$,$k$ linked)            &  &        0.57&       22926\\
Don't know                            &  &        0.46&       66810\\
\hline\hline
\end{tabular*}
}

%% file: table_reachable_panel_dim5.tex
\begin{tabular}{lcccc} \hline
 & (1) & (2) & (3) & (4) \\
VARIABLES & Correct guess & Correct guess & Correct guess & Correct guess \\ \hline
 &  &  &  &  \\
1\{Reachable j,k, Not linked j,k\} & 0.0647 & 0.0639 & 0.0549 & 0.0420 \\
 & (0.0196) & (0.0192) & (0.0194) & (0.0224) \\
 & [0.001] & [0.001] & [0.006] & [0.065] \\
1\{Reachable j,k, Linked j,k\} & 0.547 & 0.147 & 0.122 & 0.174 \\
 & (0.0227) & (0.112) & (0.108) & (0.142) \\
 & [0.000] & [0.193] & [0.261] & [0.224] \\
Avg. distance to j,k, Not linked j,k & -0.0177 & -0.0153 & -0.0115 & -0.0133 \\
 & (0.00398) & (0.00389) & (0.00400) & (0.00437) \\
 & [0.000] & [0.000] & [0.005] & [0.003] \\
Avg. distance to j,k, Linked j,k & -0.109 & -0.0943 & -0.0920 & -0.0925 \\
 & (0.00729) & (0.00656) & (0.00607) & (0.00673) \\
 & [0.000] & [0.000] & [0.000] & [0.000] \\
 &  &  &  &  \\
Observations & 66,810 & 66,810 & 66,810 & 66,810 \\
 Depvar mean & 0.199 & 0.199 & 0.199 & 0.199 \\ \hline
\end{tabular}

%% file: table_reachable_dontknow_panel_dim5.tex
\begin{tabular}{lcccc} \hline
 & (1) & (2) & (3) & (4) \\
VARIABLES & Don't know \ \ & Don't know \ \ & Don't know \ \ & Don't know \ \ \\ \hline
 &  &  &  &  \\
1\{Reachable j,k, Not linked j,k\} & -0.321 & -0.271 & -0.246 & -0.253 \\
 & (0.0377) & (0.0362) & (0.0288) & (0.0335) \\
 & [0.000] & [0.000] & [0.000] & [0.000] \\
1\{Reachable j,k, Linked j,k\} & -0.466 & -0.355 & -0.244 & -0.228 \\
 & (0.0381) & (0.153) & (0.163) & (0.216) \\
 & [0.000] & [0.023] & [0.139] & [0.293] \\
Avg. distance to j,k, Not linked j,k & 0.109 & 0.0933 & 0.0854 & 0.0811 \\
 & (0.00866) & (0.00939) & (0.00653) & (0.00686) \\
 & [0.000] & [0.000] & [0.000] & [0.000] \\
Avg. distance to j,k, Linked j,k & 0.148 & 0.123 & 0.118 & 0.122 \\
 & (0.00873) & (0.00788) & (0.00784) & (0.00887) \\
 & [0.000] & [0.000] & [0.000] & [0.000] \\
 &  &  &  &  \\
Observations & 66,810 & 66,810 & 66,810 & 66,810 \\
 Depvar mean & 0.463 & 0.463 & 0.463 & 0.463 \\ \hline
\end{tabular}

%% file: table_reachable_know_panel_dim5.tex
\begin{tabular}{lcccc} \hline
 & (1) & (2) & (3) & (4) \\
VARIABLES & Correct guess & Correct guess & Correct guess & Correct guess \\ \hline
 &  &  &  &  \\
1\{Reachable j,k, Not linked j,k\} & 0.00979 & 0.0348 & 0.0327 & -0.00996 \\
 & (0.0432) & (0.0424) & (0.0368) & (0.0484) \\
 & [0.821] & [0.414] & [0.378] & [0.837] \\
1\{Reachable j,k, Linked j,k\} & 0.534 & -0.158 & -0.101 & 0.108 \\
 & (0.0423) & (0.108) & (0.127) & (0.359) \\
 & [0.000] & [0.148] & [0.427] & [0.764] \\
Avg. distance to j,k, Not linked j,k & 0.00668 & 0.00541 & 0.00857 & 0.000630 \\
 & (0.00699) & (0.00770) & (0.00684) & (0.00720) \\
 & [0.342] & [0.484] & [0.214] & [0.930] \\
Avg. distance to j,k, Linked j,k & -0.0189 & -0.0208 & -0.0177 & -0.0110 \\
 & (0.00697) & (0.00711) & (0.00727) & (0.00924) \\
 & [0.008] & [0.005] & [0.017] & [0.239] \\
 &  &  &  &  \\
Observations & 35,909 & 35,909 & 35,909 & 35,909 \\
 Depvar mean & 0.371 & 0.371 & 0.371 & 0.371 \\ \hline
\end{tabular}

%% file: table_knowpair_panel_dim5.tex
\begin{tabular}{lcccc} \hline
 & (1) & (2) & (3) & (4) \\
VARIABLES & Know pair & Know pair & Don't know & Don't know \\ \hline
 &  &  &  &  \\
1\{Reachable j,k, Not linked j,k\} & 0.337 & 0.289 & -0.0671 & -0.0490 \\
 & (0.0393) & (0.0389) & (0.0355) & (0.0337) \\
 & [0.000] & [0.000] & [0.063] & [0.150] \\
1\{Reachable j,k, Linked j,k\} & 0.446 & 0.349 & -0.122 & -0.0548 \\
 & (0.0395) & (0.174) & (0.0338) & (0.0672) \\
 & [0.000] & [0.049] & [0.001] & [0.418] \\
Avg. distance to j,k, Not linked j,k & -0.109 & -0.0959 & 0.0323 & 0.0229 \\
 & (0.00923) & (0.00975) & (0.00736) & (0.00728) \\
 & [0.000] & [0.000] & [0.000] & [0.002] \\
Avg. distance to j,k, Linked j,k & -0.137 & -0.116 & 0.0447 & 0.0349 \\
 & (0.00948) & (0.00834) & (0.00641) & (0.00738) \\
 & [0.000] & [0.000] & [0.000] & [0.000] \\
 &  &  &  &  \\
Observations & 22,270 & 22,270 & 42,750 & 42,750 \\
Know pair only & $ $ & $ $ & \checkmark & \checkmark \\
Depvar mean & 0.640 & 0.640 & 0.160 & 0.160 \\
 Number of (\textit{i, jk})s &  &  & 14250 & 14250 \\ \hline
\end{tabular}

%% file: table_lasso_panel_dim5.tex
\begin{tabular}{lcc} \hline
 & (1) & (2) \\
VARIABLES & Correct guess & Don't know \\ \hline
 &  &  \\
Avg. distance to j,k, Not linked j,k & -0.0269 & 0.105 \\
 & (0.00367) & (0.00603) \\
 & [0.000] & [0.000] \\
Avg. distance to j,k, Linked j,k &  & 0.0899 \\
 &  & (0.00554) \\
 &  & [0.000] \\
Centrality (std.) of i, Linked j,k & 0.0134 & -0.0125 \\
 & (0.00582) & (0.00560) \\
 & [0.024] & [0.029] \\
Avg. centrality (std.) of j,k, Not linked j,k &  & -0.00211 \\
 &  & (0.00692) \\
 &  & [0.761] \\
Avg. centrality (std.) of j,k, Linked j,k & 0.0121 &  \\
 & (0.00520) &  \\
 & [0.022] &  \\
 &  &  \\
Observations & 57,597 & 63,558 \\
Reachable pairs only & \checkmark & \checkmark \\
LASSO-selected Demographic controls & \checkmark & \checkmark \\
LASSO-selected Village FE & \checkmark & \checkmark \\
 Depvar mean & 0.204 & 0.459 \\ \hline
\end{tabular}

%% file: ProofAppendix.tex
\subsection*{Proof of Proposition \ref{proposition:learning}}

\paragraph{\textit{Proof of part (a)}}
\vskip 0.05in

Throughout, let $\omega^n = (\omega_1,\dots, \omega_n)$, where $\omega_i$ denotes the point estimates of agent $i$. Without loss of generality, suppose the true underlying state and the realized network are $\theta$ and $\g^2$, respectively. By assumption, the realized network is observable to agent $0$. Consequently, agent $0$'s posterior likelihood ratio for an arbitrary pair of states $\hat\theta$ and $\tilde\theta$ given her observations is equal to 
\begin{align}
\frac{f(\hat\theta|s_0, \omega^n, \g^2)}{f(\tilde{\theta}|s_0, \omega^n,\g^2)} = \frac{ f(s_0, \omega^n|\hat\theta, \g^2) }{f(s_0, \omega^n|\tilde\theta, \g^2) } = \frac{f(s_0|\hat\theta) f(\omega^n|\hat\theta, \g^2) }{f(s_0|\tilde\theta)f(\omega^n|\tilde\theta, \g^2) },\label{eq:llr}
\end{align}
where we are using the assumption that agents have a common uniform prior beliefs over the real line and the fact that signal $s_0$ is conditionally independent from the point estimates of all other agents. On the other hand, since agents $j\in \{3,\dots, n\}$ can observe agent 2's point estimate $\omega_2 = s_2$, it is immediate that $\omega_j = (s_j +s_2)/2$. Therefore, for any arbitrary state $\tilde\theta\in\mathbb{R}$, 
\begin{align*}
f(\omega^n| \tilde\theta, \g^2) & = \frac{1}{\left(\sqrt{2\pi} \sigma\right)^{n}} \exp\left(-\frac{1}{2\sigma^2} \left((\tilde\theta-\omega_1)^2 + (\tilde\theta-\omega_2)^2 + \sum_{i=3}^n (\tilde\theta - 2\omega_i + \omega_2)^2\right)\right ).
\end{align*}
The juxtaposition of the above equation with equation \eqref{eq:llr} therefore implies that agent $0$'s posterior log-likelihood ratio for any given pair of states $\hat\theta$ and $\tilde\theta$ is equal to
\begin{align} \label{eq:llg1}
\log \frac{f(\hat\theta|s_0, \omega^n, \g^2)}{f(\tilde{\theta}|s_0, \omega^n,\g^2)} & = \frac{1}{2\sigma^2} (\tilde\theta - \hat\theta) \left((n+1)(\hat\theta + \tilde\theta) - 2s_0- 2\omega_1 - 2\omega_2 - 2\sum_{i=3}^n \left( 2\omega_i - \omega_2\right) \right).
\end{align}
Set $\hat\theta = \theta$, where $\theta$ is the underlying state of the world. Multiplying both sides of the above equation by $1/n$ and taking the limit as $n\to\infty$ implies that 
\begin{align*}
\lim_{n\to\infty} \frac{1}{n}\log \frac{f(\theta|s_0, \omega^n, \g^2)}{f(\tilde{\theta}|s_0, \omega^n,\g^2)}& = \frac{1}{2\sigma^2} (\tilde\theta - \theta)^2,
\end{align*}
where we are using the fact that by the law of large numbers, $\lim_{n\to\infty }(1/n)\sum_{i=3}^n \omega_i   = (\omega_2 +  \theta)/2$ with probability 1. Therefore, for any $\tilde\theta \neq \theta$, 
\begin{align*}
\lim_{n\to\infty} \frac{f(\theta|s_0, \omega^n, \g^2)}{f(\tilde{\theta}|s_0, \omega^n,\g^2)}& = \infty
\end{align*}
with probability one. Consequently, agent $0$ assigns an asymptotic belief of one on the true state $\theta$ as $n\to\infty$ almost surely.
\qed\newline

\paragraph{\textit{Proof of part (b)}}

Without loss of generality let $\theta$ and $\g^1$ denote the realized state and network, respectively. However, suppose that agent $0$ has misspecified beliefs about the network by mistakenly believing that the underlying network is $\g^2$. Following an argument identical to the proof of part (a), the log-likelihood ratio of agent $0$'s posterior beliefs is given by \eqref{eq:llg1}. Consequently, the log-likelihood ratio of agent $0$'s posterior beliefs as $n\to\infty$ satisfies
\begin{align}\label{eq:FML}
\lim_{n\to\infty} \frac{1}{n}\log \frac{f(\hat\theta|s_0, \omega^n, \g^2)}{f(\tilde{\theta}|s_0, \omega^n,\g^2)}& = \frac{1}{2\sigma^2} (\tilde\theta - \hat\theta)(\hat\theta + \tilde\theta - 2(\theta + \omega_1 - \omega_2))
\end{align}
almost surely. In the above expression, we are using the fact that the true underlying network is $\g^1$, and hence, $\lim_{n\to\infty }(1/n)\sum_{i=3}^n \omega_i   = (\omega_1 +  \theta)/2$ with probability one. Setting $\hat\theta = \theta + \omega_1 - \omega_2$ therefore implies that the right-hand side of \eqref{eq:FML} is strictly positive for all $\tilde\theta \neq \theta + \omega_1 - \omega_2$. Consequently, 
\begin{align}\label{eq:aux}
\lim_{n\to\infty} \frac{f(\hat\theta|s_0, \omega^n, \g^2)}{f(\tilde{\theta}|s_0, \omega^n,\g^2)}& = \infty
\end{align}
almost surely for all $\tilde\theta \neq \theta + \omega_1 - \omega_2$. In other words, as $n\to\infty$, agent $0$ becomes certain that the realized state is $\theta + \omega_1 - \omega_2$; a state that is distinct from the true underlying state $\theta$ with probability one. 
\qed\newline

\paragraph{\textit{Proof of part (c)}}
Without loss of generality let $\theta$ and $\g^1$ denote the realized state and network, respectively, and let $\hat\theta \neq\theta$. Agent $0$'s posterior likelihood ratios are given by
\begin{align*}
 \frac{f(\theta|s_0,\omega^n)}{f(\hat{\theta}|s_0,\omega^n)}  & = \frac{\mathbb{P}(\g^1) f(s_0,\omega^n|\theta, \g^1) + \mathbb{P}( \g^2)f(s_0,\omega^n|\theta,  \g^2)}{\mathbb{P}(\g^1) f(s_0,\omega^n|\hat\theta, \g^1) + \mathbb{P}( \g^2)f(s_0,\omega^n|\hat\theta, \g^2)},
\end{align*} 
where we are using Bayes' rule and the assumption that agents have improper uniform prior beliefs. Consequently, 
\begin{align*}
 \frac{f(\theta|s_0,\omega^n)}{f(\hat{\theta}|s_0,\omega^n)}  & \leq \frac{\mathbb{P}(\g^1) f(s_0,\omega^n|\theta, \g^1)}{ \mathbb{P}( \g^2)f(s_0,\omega^n|\hat\theta, \g^2)} + \frac{f(s_0,\omega^n|\theta,  \g^2)}{f(s_0,\omega^n|\hat\theta, \g^2)} = \frac{\mathbb{P}(\g^1) f(s_0|\theta)f(\omega^n|\theta, \g^1)}{ \mathbb{P}( \g^2)f(s_0|\hat\theta)f(\omega^n|\hat\theta, \g^2)} + \frac{f(\theta|s_0,\omega^n,  \g^2)}{f(\hat\theta|s_0,\omega^n, \g^2)}.
\end{align*}
Let $\hat{\theta} = \theta + \omega_1  -\omega_2$. Equation \eqref{eq:aux} in the proof of part (b) implies that the last term on the right-hand side of the above equation converges to zero as $n\to\infty$ almost surely whenever the true state-network pair is $(\theta,\g^1)$. Therefore, 
\begin{align}\label{eq:limit}
\limsup_{n\to\infty} \frac{f(\theta|s_0,\omega^n)}{f(\hat{\theta}|s_0,\omega^n)}  & \leq \frac{\mathbb{P}(\g^1)}{\mathbb{P}( \g^2)} \frac{f(s_0|\theta)}{f(s_0|\hat\theta)}\limsup_{n\to\infty}  \frac{f(\omega^n|\theta, \g^1)}{ f(\omega^n|\hat\theta, \g^2)}.
 \end{align}
On the other hand, note that 
\begin{align*}
\log\frac{f(\omega^n|\theta, \g^1)}{ f(\omega^n|\hat\theta, \g^2)} & =  \frac{1}{2\sigma^2}\left((\hat\theta-\omega_1)^2 - (\theta-\omega_2)^2\right)  +  \frac{1}{2\sigma^2} \left((\hat\theta-{\omega}_2)^2 - (\theta-{\omega}_1)^2\right)\\
&  \quad + \frac{1}{2\sigma^2} \sum_{i=3}^n \Big((\hat\theta - 2\omega_i + \omega_2)^2 - (\theta - 2\omega_i + \omega_1)^2 \Big) .
\end{align*}
Since $\hat{\theta} = \theta + \omega_1  -\omega_2$, the first and third terms on the right-hand side of the above equation is equal to zero. As a result,
\begin{align*}
\log\frac{f(\omega^n|\theta, \g^1)}{ f(\omega^n|\hat\theta, \g^2)} & ={2} (\omega_1-\omega_2)(\theta -\omega_2)/\sigma^2
\end{align*}
which is a finite constant that does not depend on $n$. Hence, the inequality in \eqref{eq:limit} reduces to 
\begin{align*}
\limsup_{n\to\infty} \frac{f(\theta|s_0,\omega^n)}{f(\hat{\theta}|s_0,\omega^n)}  & < \infty
\end{align*}
almost surely. Hence, the belief that agent $0$ assigns to the true state $\theta$ remains bounded away from 1. This observation, coupled with the fact that a Bayesian agent never rules out the true state as impossible implies that agent $0$ remains uncertain forever with probability one. 
\qed


\subsection*{Proof of Proposition \ref{proposition:BCZ}}

$\quad$\\
We start by analyzing the complete information benchmark. Under complete information about the realized network, the first-order condition of agent $i$ is given by 
\begin{align*}
x_i & = \theta_i + \alpha \sum_{j\neq i} g_{ij}x_j. 
\end{align*}
Therefore, the vector of equilibrium actions in the complete information variant of the game is equal to $x^{\mathrm{com}} = (I-\alpha \g)^{-1}\theta$, thus establishing \eqref{eq:complete}. Note that the assumptions that $\sum_{j\neq i} \max \supp(f_{ij}) \leq 1$ and $\alpha<1$ guarantee that $(I-\alpha\g)^{-1}$ always exists and is element-wise non-negative. 

Next, consider the incomplete information game, in which agent $i$ only observes the realization of the weights in the set $\{g_{ij}: j\neq i\}$, with no information about $g_{jk}$ for $j\neq i$. We establish \eqref{eq:incomplete} by verifying that the vector $x = (I+\alpha \g L)\theta$ satisfies the best-response equations of all agents simultaneously, where $L =  (I-\alpha \mathbb{E}[\g])^{-1}$. To this end, recall that the first-order condition of agent $i$ is given by 
\begin{align}\label{eq:FOC}
x_i & = \theta_i + \alpha \sum_{j\neq i} g_{ij}\mathbb{E}_i[x_j],
\end{align}
where $\mathbb{E}_i[x_j]$ is $i$'s expectation of $j$'s action conditional on observing her own local neighborhood weights $g_{ij}$. Suppose that the action of each agent $j\neq i$ is given by the conjectured actions $x_j = \theta_j + \alpha \sum_{k\neq j}  \sum_{r=1}^n g_{jk}{\ell}_{kr} \theta_r$. Plugging this expression into the left-hand side of agent $i$'s first-order condition in \eqref{eq:FOC} implies that 
\begin{align*}
x_i & = \theta_i + \alpha \sum_{j\neq i} g_{ij} \theta_j + \alpha^2 \sum_{j\neq i} \sum_{k\neq j} \sum_{r=1}^n g_{ij}  \mathbb{E}[g_{jk}]\ell_{kr}\theta_r,
\end{align*}
where we are using the fact that $\mathbb{E}_i[g_{jk}] = \mathbb{E}[g_{jk}]$. This is a consequence of the assumption that the information set of agent $i$ provides no information about $g_{jk}$ when $j\neq i$. Furthermore, the fact that $L = \sum_{m=0}^\infty \alpha^m \mathbb{E}^m[\g]$ implies that $\alpha \mathbb{E}[\g] L = L-I$. Therefore, it is immediate that 
\begin{align*}
x_i & = \theta_i + \alpha \sum_{j\neq i} \sum_{r=1}^n g_{ij}\ell_{jr}\theta_r.
\end{align*}
Consequently, the strategy profile $x = (I+\alpha \g L)\theta$ satisfies the first-order conditions of all agents simultaneously, establishing \eqref{eq:incomplete}.

We complete the proof of Proposition \ref{proposition:BCZ} by establishing that $\mathbb{E}[x_i^{\mathrm{inc}}] < \mathbb{E}[x_i^{\mathrm{com}}]$ for all agents $i$. As a first observation, note that equation \eqref{eq:incomplete} implies the ex ante vector of equilibrium actions in the incomplete information game is given by $\mathbb{E}[x^{\mathrm{inc}}] = \mathbb{E}[I+\alpha \g (I-\alpha \mathbb{E}[\g])^{-1}]\theta$. Consequently, 
\begin{align}\label{eq:eqinc}
\mathbb{E}[x^{\mathrm{inc}}] & = \sum_{k=0}^\infty \alpha^k \mathbb{E}^k[\g]\theta. 
\end{align}
On the other hand, equation \eqref{eq:complete} implies that the ex ante vector of equilibrium actions of actions in the complete information benchmark is given by 
\begin{align}\label{eq:eqcom}
\mathbb{E}[x^{\mathrm{com}}] & = \mathbb{E}[(I-\alpha \g)^{-1}]\theta = \sum_{k=0}^\infty \alpha^k \mathbb{E}[\g^k]\theta. 
\end{align} 
The juxtaposition of \eqref{eq:eqinc} and \eqref{eq:eqcom} implies that the proof is complete once we show that $\mathbb{E}[\g^k] \geq \mathbb{E}^k[\g]$ for all non-negative integers $k$, with at least one inequality being strict. We show this by establishing that $\mathbb{E}[g_{j_1,j_2}g_{j_2,j_3}\dots g_{j_{k-1},j_k}] \geq \mathbb{E}[g_{j_1,j_2}]\dots \mathbb{E}[g_{j_{k-1},j_k}]$ for any sequence of agents $j_1,j_2,\dots, j_k$ and all $k$. 

To prove the this claim, note that if $j_m = j_{m+1}$, then both sides of the above inequality are equal to zero. Therefore, suppose that the sequence of agents $j_1,j_2,\dots, j_k$ is such that $j_m\neq j_{m+1}$ throughout. We have,
\begin{align*}
\mathbb{E}[g_{j_1,j_2}g_{j_2,j_3}\dots g_{j_{k-1},j_k}]  & = \mathbb{E}\left[\prod_{i=1}^n \prod_{l\neq i} \prod_{r\in B(i,l)} g_{j_r,j_{r+1}}\right] =\prod_{i=1}^n \prod_{l\neq i}\mathbb{E}\left[ \prod_{r\in B(i,l)} g_{j_r,j_{r+1}}\right]
\end{align*}
where $B(i,l) = \{r: j_r =i, j_{r+1}=l\}$ and the second equality is a consequence of the assumption that all elements of $\g$ are drawn independently from one another. Let $b_{il} = |B(i,l)|$. We have,  
\begin{align*}
\mathbb{E}[g_{j_1,j_2}g_{j_2,j_3}\dots g_{j_{k-1},j_k}]  &  = \prod_{i=1}^n \prod_{l\neq i} \mathbb{E}[ g_{il}^{b_{il}}] \geq  \prod_{i=1}^n \prod_{l\neq i} \mathbb{E}[ g_{il}]^{b_{il}},
\end{align*}
where the second implication is a consequence of Jensen's inequality. Furthermore, note that the inequality is strict whenever there exists pair of agents $i$ and $l$ in the sequence such that $b_{il}\geq 2$. As a result, 
\begin{align*}
\mathbb{E}[g_{j_1,j_2}g_{j_2,j_3}\dots g_{j_{k-1},j_k}]  &  \geq \prod_{i=1}^n \prod_{l\neq i} \prod_{r\in B(i,l)} \mathbb{E}[g_{j_r,j_{r+1}}] = \mathbb{E}[g_{j_1,j_2}]\mathbb{E}[g_{j_2,j_3}]\dots \mathbb{E}[g_{j_{k-1},j_k}].
\end{align*}
The above inequality guarantees $\mathbb{E}[\g^k] \geq \mathbb{E}^k[\g]$ for all integer $k\geq 0$ and hence completes the proof.
\qed


\subsection*{Proof of Proposition \ref{proposition:formation}}

%
%

Consider a monotone symmetric equilibrium of the incomplete information game, according to which each agent $i$ with attribute  $z_i \in \{a,b\}$ chooses to link to an agent $j$ with attribute $z_j$ if and only if $\epsilon_{ij}$ falls below the threshold $\tau_{z_i z_j}$. Hence, any monotone symmetric equilibrium is characterized by four thresholds $\tau_{aa}$, $\tau_{ab}$, $\tau_{ba}$, and $\tau_{bb}$. 

As a starting point, we determine the moments 
 of these thresholds that are identifiable from the observed network $\g$. Recall that the formation of each link requires the mutual consent of both parties. This means that a link between agents $i$ and $j$ with attributes $z_i = z_j = a$ exists if and only if $\epsilon_{ij},\epsilon_{ji} \leq \tau_{aa}$. Consequently, 
\begin{align}
p_{aa} & = F^2(\tau_{aa}),\label{eq:paa}
\end{align}
where $p_{aa}$ is the fraction of linkages within type-$a$ agents and $F(\cdot)$ denotes the common distribution of preference shocks. A similar argument implies that
\begin{align}
p_{bb} & = F^2(\tau_{bb}) \label{eq:pbb}\\
p_{ab} & = F(\tau_{ab})F(\tau_{ba}),\label{eq:pab}
\end{align}
where $p_{bb}$ and $p_{ab}$ denote the fraction of linkages within type-$b$ agents and across agents of types $a$ and $b$, respectively. 

Next we use the equilibrium conditions to relate the model's structural parameters $\alpha$, $\beta$, and $\gamma$ to the observables. More specifically, note that for $\tau_{aa}$ to correspond to an equilibrium threshold, agent $i$ with attribute $z_i = a$ must be indifferent between linking to an agent $j$ with the same attribute whenever $\epsilon_{ij} = \tau_{aa}$. In other words, 
\begin{align*}
\tau_{aa} & = \alpha + \beta + \frac{\gamma}{n-1}\sum_{k\neq i,j}\mathbb{E}[g_{jk}].
\end{align*}
Furthermore, note that $\mathbb{E}[g_{jk}] = F^2(\tau_{aa})$ if $z_j = z_k = a$, whereas $\mathbb{E}[g_{jk}] = F(\tau_{ab})F(\tau_{ba})$ if $z_j = a$ and $z_k = b$. Replacing for these terms from equations \eqref{eq:paa} and \eqref{eq:pab} therefore implies that 
\begin{align}\label{eq:identification1}
F^{-1}(\sqrt{p_{aa}}) & = \alpha + \beta + \frac{\gamma}{n-1}\Big((n\lambda -2)p_{aa} + n(1-\lambda)p_{ab} \Big),
\end{align}
where $\lambda$ denotes the fraction of agents with attribute $a$. A similar argument applied to linkages within type-$b$ agents implies that 
\begin{align}\label{eq:identification2}
F^{-1}(\sqrt{p_{bb}}) & = \alpha + \beta +  \frac{\gamma}{n-1}\Big(n\lambda p_{ab} + (n(1-\lambda)-2) p_{bb} \Big).
\end{align}
Equations \eqref{eq:identification1} and \eqref{eq:identification2} provide the econometrician with two equations that relate network observables to the structural parameters $\alpha$, $\beta$, and $\gamma$ and hence can be used for identification. Specifically, as long as 
\begin{align}\label{eq:lambda}
\lambda \neq \frac{(n-2)(p_{bb}-p_{ab}) + 2(p_{aa}-p_{ab})}{n(p_{aa} + p_{bb}-2p_{ab})},
\end{align}
the two equations are linearly independent and hence $\gamma$ and $\alpha+\beta$ are point-identified. 

With the above in hand, the next step is to determine conditions under which $\alpha$ and $\beta$ are separably identifiable. To this end, note that $\tau_{ab}$ and $\tau_{ba}$ correspond to equilibrium thresholds if and only if 
\begin{align*}
\tau_{ab} & = \alpha  + \frac{\gamma}{n-1}\Big((n\lambda -1)p_{ab} + (n(1-\lambda)-1)p_{bb} \Big)\\
\tau_{ba} & = \alpha + \frac{\gamma}{n-1}\Big((n\lambda-1)p_{aa} + (n(1-\lambda) -1)p_{ab} \Big),
\end{align*}
where once again we are using equations \eqref{eq:paa}--\eqref{eq:pab} to replace for the thresholds in terms of the observables fractions $p_{aa}$, $p_{ab}$, and $p_{bb}$. Consequently, 
\begin{align*}
\tau_{ab} & = F^{-1}(\sqrt{p_{bb}}) - \beta + \gamma(p_{bb} - p_{ab}) /(n-1)\\
\tau_{ba} & = F^{-1}(\sqrt{p_{aa}}) - \beta + \gamma (p_{aa} - p_{ab}) /(n-1).
\end{align*}
Using equation \eqref{eq:pab} one more time implies that $G(\beta,\gamma) = p_{ab}$, where 
\begin{align*}
G(\beta,\gamma) & = F\!\left(F^{-1}(\sqrt{p_{aa}}) - \beta + \gamma \frac{p_{aa} - p_{ab}}{n-1}\right) \cdot F\!\left(F^{-1}(\sqrt{p_{bb}}) - \beta + \gamma \frac{p_{bb} - p_{ab}}{n-1}\right).
\end{align*}
Recall that, by assumption, the preference shocks have full support over $\mathbb{R}$, which implies that $F$ is strictly increasing. This means that $G(\beta,\gamma)$ is strictly decreasing over its domain. This observation, alongside the fact that $\lim_{\beta\to\infty} G(\beta,\gamma) = 0$ and $\lim_{\beta\to-\infty} G(\beta,\gamma) = 1$, implies that for every $\gamma$, there exists a unique $\beta$ such that $G(\beta,\gamma) = p_{ab}$. This implying that $\beta$ is also point-identified. Since $\alpha + \beta$ was already identified, this means that all structural parameters are point-identified for all values of $n$. 

The proof is complete once we verify that, as $n\to\infty$, the identification requirement on $\lambda$ expressed in \eqref{eq:lambda} is always satisfied as long as $\lambda\neq 1/2$. To this end, suppose that \eqref{eq:lambda} is violated, that is,
\begin{align*}
\lambda = \frac{(n-2)(p_{bb}-p_{ab}) + 2(p_{aa}-p_{ab})}{n(p_{aa} + p_{bb}-2p_{ab})},
\end{align*}
which reduces to
\begin{align}\label{eq:lackof}
\lambda = \frac{p_{bb}-p_{ab}}{p_{aa} + p_{bb}-2p_{ab}}
\end{align}
as $n\to\infty$. Plugging back this expression into equations \eqref{eq:identification1} and \eqref{eq:identification2} implies that \eqref{eq:lambda} is violated asymptotically only if $p_{aa} = p_{bb}$, which by \eqref{eq:lackof}, is equivalent to $\lambda = 1/2$. Thus, as long as $\lambda \neq 1/2$, the identification requirement \eqref{eq:lambda} on $\lambda$ is always satisfied as $n\to\infty$.
\qed

%% file: apdx_w_cent.tex
\begin{table}[!h]
\centering
\caption{Network Knowledge as a function of Distance and Centralities \label{tab:reachable}}
\scalebox{0.8}{\begin{threeparttable}
\input{table_reachable_panel_dim5_cent.tex}
\begin{tablenotes}
Notes: Standard errors (clustered at village level) are reported in parentheses. $p$-values are reported in brackets. Demographic controls include average geographic distance of $i$ from $j$ and $k$ and dummies for each of whether $i$ is of the same caste, subcaste, has the same electrification status, has the same roof type, has the same number of rooms, and has the same house ownership status for each of $j$ and $k$, when $g_{jk}$ is 1 and 0. They also include the above-mentioned dummies for when $j$ and $k$ have the same demographic traits. 
\end{tablenotes}
\end{threeparttable}}
\end{table}

\clearpage

\begin{table}[!h]
\centering
\caption{Don't Know as a function of Distance and Centralities \label{tab:reachable_dontknow}}
\scalebox{1}{\begin{threeparttable}
\input{table_reachable_dontknow_panel_dim5_cent.tex}
\begin{tablenotes}
Notes: Standard errors (clustered at village level) are reported in parentheses. $p$-values are reported in brackets. Demographic controls include average geographic distance of $i$ from $j$ and $k$ and dummies for each of whether $i$ is of the same caste, subcaste, has the same electrification status, has the same roof type, has the same number of rooms, and has the same house ownership status for each of $j$ and $k$, when $g_{jk}$ is 1 and 0. They also include the above-mentioned dummies for when $j$ and $k$ have the same demographic traits. 
\end{tablenotes}
\end{threeparttable}}
\end{table}

\clearpage

\begin{table}[!h]
\centering
\caption{Correct Guess, conditional on sure guesses only, as a function of distance and centralities \label{tab:reachable_know}}
\scalebox{1}{\begin{threeparttable}
\input{table_reachable_know_panel_dim5_cent.tex}
\begin{tablenotes}
Notes: Standard errors (clustered at village level) are reported in parentheses. $p$-values are reported in brackets. Demographic controls include average geographic distance of $i$ from $j$ and $k$ and dummies for each of whether $i$ is of the same caste, subcaste, has the same electrification status, has the same roof type, has the same number of rooms, and has the same house ownership status for each of $j$ and $k$, when $g_{jk}$ is 1 and 0. They also include the above-mentioned dummies for when $j$ and $k$ have the same demographic traits. 
\end{tablenotes}
\end{threeparttable}}
\end{table}

\clearpage

\begin{table}[!h]
\centering
\caption{Network Knowledge, conditional on knowing $j$,$k$, as a function of Distance and Centralities \label{tab:knowpair}}
\scalebox{1}{\begin{threeparttable}
\input{table_knowpair_panel_dim5_cent.tex}
\begin{tablenotes}
Notes: Standard errors (clustered at village level) are reported in parentheses. $p$-values are reported in brackets. Demographic controls include average geographic distance of $i$ from $j$ and $k$ and dummies for each of whether $i$ is of the same caste, subcaste, has the same electrification status, has the same roof type, has the same number of rooms, and has the same house ownership status for each of $j$ and $k$, when $g_{jk}$ is 1 and 0. They also include the above-mentioned dummies for when $j$ and $k$ have the same demographic traits.
\end{tablenotes}
\end{threeparttable}}
\end{table}

\clearpage

%% file: table_reachable_panel_dim5_cent.tex
\begin{tabular}{lcccc} \hline
 & (1) & (2) & (3) & (4) \\
VARIABLES & Correct guess & Correct guess & Correct guess & Correct guess \\ \hline
 &  &  &  &  \\
1\{Reachable j,k, Not linked j,k\} & 0.0647 & 0.0639 & 0.0549 & 0.0420 \\
 & (0.0196) & (0.0192) & (0.0194) & (0.0224) \\
 & [0.001] & [0.001] & [0.006] & [0.065] \\
1\{Reachable j,k, Linked j,k\} & 0.547 & 0.147 & 0.122 & 0.174 \\
 & (0.0227) & (0.112) & (0.108) & (0.142) \\
 & [0.000] & [0.193] & [0.261] & [0.224] \\
Avg. distance to j,k, Not linked j,k & -0.0177 & -0.0153 & -0.0115 & -0.0133 \\
 & (0.00398) & (0.00389) & (0.00400) & (0.00437) \\
 & [0.000] & [0.000] & [0.005] & [0.003] \\
Avg. distance to j,k, Linked j,k & -0.109 & -0.0943 & -0.0920 & -0.0925 \\
 & (0.00729) & (0.00656) & (0.00607) & (0.00673) \\
 & [0.000] & [0.000] & [0.000] & [0.000] \\
Centrality (std.) of i, Not linked j,k & 0.00265 & 0.00224 & 0.00325 &  \\
 & (0.00285) & (0.00286) & (0.00280) &  \\
 & [0.356] & [0.437] & [0.250] &  \\
Centrality (std.) of i, Linked j,k & 5.61e-05 & 0.00152 & 0.00247 & -6.32e-06 \\
 & (0.00521) & (0.00520) & (0.00510) & (0.00588) \\
 & [0.991] & [0.772] & [0.629] & [0.999] \\
Avg. centrality (std.) of j,k, Not linked j,k & -0.000971 & -0.000282 & 0.00263 & -0.000425 \\
 & (0.00343) & (0.00328) & (0.00319) & (0.00349) \\
 & [0.778] & [0.932] & [0.412] & [0.903] \\
Avg. centrality (std.) of j,k, Linked j,k & -0.00820 & -0.00782 & -0.00704 & -0.00619 \\
 & (0.00513) & (0.00521) & (0.00505) & (0.00524) \\
 & [0.114] & [0.138] & [0.167] & [0.241] \\
 &  &  &  &  \\
Observations & 66,810 & 66,810 & 66,810 & 66,810 \\
Dimension FE & \checkmark & \checkmark & \checkmark & \checkmark \\
Village FE & $ $ & $ $ & \checkmark & $ $ \\
Respondent FE & $ $ & $ $ & $ $ & \checkmark \\
Demographic controls & $ $ & \checkmark & \checkmark & \checkmark \\
 Depvar mean & 0.199 & 0.199 & 0.199 & 0.199 \\ \hline
\end{tabular}

%% file: table_reachable_dontknow_panel_dim5_cent.tex
\begin{tabular}{lcccc} \hline
 & (1) & (2) & (3) & (4) \\
VARIABLES & Don't know & Don't know & Don't know & Don't know \\ \hline
 &  &  &  &  \\
1\{Reachable j,k, Not linked j,k\} & -0.321 & -0.271 & -0.246 & -0.253 \\
 & (0.0377) & (0.0362) & (0.0288) & (0.0335) \\
 & [0.000] & [0.000] & [0.000] & [0.000] \\
1\{Reachable j,k, Linked j,k\} & -0.466 & -0.355 & -0.244 & -0.228 \\
 & (0.0381) & (0.153) & (0.163) & (0.216) \\
 & [0.000] & [0.023] & [0.139] & [0.293] \\
Avg. distance to j,k, Not linked j,k & 0.109 & 0.0933 & 0.0854 & 0.0811 \\
 & (0.00866) & (0.00939) & (0.00653) & (0.00686) \\
 & [0.000] & [0.000] & [0.000] & [0.000] \\
Avg. distance to j,k, Linked j,k & 0.148 & 0.123 & 0.118 & 0.122 \\
 & (0.00873) & (0.00788) & (0.00784) & (0.00887) \\
 & [0.000] & [0.000] & [0.000] & [0.000] \\
Centrality (std.) of i, Not linked j,k & 0.00201 & 0.00139 & -0.00383 &  \\
 & (0.00608) & (0.00617) & (0.00531) &  \\
 & [0.742] & [0.822] & [0.472] &  \\
Centrality (std.) of i, Linked j,k & -0.00335 & -0.00565 & -0.00979 & -0.00396 \\
 & (0.00595) & (0.00587) & (0.00571) & (0.00649) \\
 & [0.574] & [0.339] & [0.091] & [0.543] \\
Avg. centrality (std.) of j,k, Not linked j,k & -0.00535 & -0.00438 & -0.00927 & -0.00413 \\
 & (0.00761) & (0.00740) & (0.00675) & (0.00641) \\
 & [0.484] & [0.556] & [0.174] & [0.522] \\
Avg. centrality (std.) of j,k, Linked j,k & 0.00241 & 0.00109 & 0.000736 & 0.00668 \\
 & (0.00596) & (0.00614) & (0.00604) & (0.00650) \\
 & [0.687] & [0.859] & [0.903] & [0.307] \\
 &  &  &  &  \\
Observations & 66,810 & 66,810 & 66,810 & 66,810 \\
Dimension FE & \checkmark & \checkmark & \checkmark & \checkmark \\
Village FE & $ $ & $ $ & \checkmark & $ $ \\
Respondent FE & $ $ & $ $ & $ $ & \checkmark \\
Demographic controls & $ $ & \checkmark & \checkmark & \checkmark \\
 Depvar mean & 0.463 & 0.463 & 0.463 & 0.463 \\ \hline
\end{tabular}

%% file: table_reachable_know_panel_dim5_cent.tex
\begin{tabular}{lcccc} \hline
 & (1) & (2) & (3) & (4) \\
VARIABLES & Correct guess & Correct guess & Correct guess & Correct guess \\ \hline
 &  &  &  &  \\
1\{Reachable j,k, Not linked j,k\} & 0.00979 & 0.0348 & 0.0327 & -0.00996 \\
 & (0.0432) & (0.0424) & (0.0368) & (0.0484) \\
 & [0.821] & [0.414] & [0.378] & [0.837] \\
1\{Reachable j,k, Linked j,k\} & 0.534 & -0.158 & -0.101 & 0.108 \\
 & (0.0423) & (0.108) & (0.127) & (0.359) \\
 & [0.000] & [0.148] & [0.427] & [0.764] \\
Avg. distance to j,k, Not linked j,k & 0.00668 & 0.00541 & 0.00857 & 0.000630 \\
 & (0.00699) & (0.00770) & (0.00684) & (0.00720) \\
 & [0.342] & [0.484] & [0.214] & [0.930] \\
Avg. distance to j,k, Linked j,k & -0.0189 & -0.0208 & -0.0177 & -0.0110 \\
 & (0.00697) & (0.00711) & (0.00727) & (0.00924) \\
 & [0.008] & [0.005] & [0.017] & [0.239] \\
Centrality (std.) of i, Not linked j,k & 0.00605 & 0.00496 & 0.00213 &  \\
 & (0.00530) & (0.00533) & (0.00526) &  \\
 & [0.257] & [0.355] & [0.686] &  \\
Centrality (std.) of i, Linked j,k & -0.00265 & -0.00169 & -0.00366 & 0.00105 \\
 & (0.00510) & (0.00523) & (0.00520) & (0.00851) \\
 & [0.605] & [0.747] & [0.484] & [0.902] \\
Avg. centrality (std.) of j,k, Not linked j,k & -0.00440 & -0.00280 & -0.00276 & -0.00474 \\
 & (0.00605) & (0.00608) & (0.00544) & (0.00666) \\
 & [0.470] & [0.647] & [0.613] & [0.479] \\
Avg. centrality (std.) of j,k, Linked j,k & -0.00999 & -0.0111 & -0.0103 & -0.00679 \\
 & (0.00587) & (0.00603) & (0.00631) & (0.00768) \\
 & [0.093] & [0.070] & [0.105] & [0.379] \\
 &  &  &  &  \\
Observations & 35,909 & 35,909 & 35,909 & 35,909 \\
Sure guess only & \checkmark & \checkmark & \checkmark & \checkmark \\
Dimension FE & \checkmark & \checkmark & \checkmark & \checkmark \\
Village FE & $ $ & $ $ & \checkmark & $ $ \\
Respondent FE & $ $ & $ $ & $ $ & \checkmark \\
Demographic controls & $ $ & \checkmark & \checkmark & \checkmark \\
 Depvar mean & 0.371 & 0.371 & 0.371 & 0.371 \\ \hline
\end{tabular}

%% file: table_knowpair_panel_dim5_cent.tex
\begin{tabular}{lcccc} \hline
 & (1) & (2) & (3) & (4) \\
VARIABLES & Know pair & Know pair & Don't know & Don't know \\ \hline
 &  &  &  &  \\
1\{Reachable j,k, Not linked j,k\} & 0.337 & 0.289 & -0.0671 & -0.0490 \\
 & (0.0393) & (0.0389) & (0.0355) & (0.0337) \\
 & [0.000] & [0.000] & [0.063] & [0.150] \\
1\{Reachable j,k, Linked j,k\} & 0.446 & 0.349 & -0.122 & -0.0548 \\
 & (0.0395) & (0.174) & (0.0338) & (0.0672) \\
 & [0.000] & [0.049] & [0.001] & [0.418] \\
Avg. distance to j,k, Not linked j,k & -0.109 & -0.0959 & 0.0323 & 0.0229 \\
 & (0.00923) & (0.00975) & (0.00736) & (0.00728) \\
 & [0.000] & [0.000] & [0.000] & [0.002] \\
Avg. distance to j,k, Linked j,k & -0.137 & -0.116 & 0.0447 & 0.0349 \\
 & (0.00948) & (0.00834) & (0.00641) & (0.00738) \\
 & [0.000] & [0.000] & [0.000] & [0.000] \\
Centrality (std.) of i, Not linked j,k & 0.000393 & 0.000918 & 0.00344 & 0.00270 \\
 & (0.00531) & (0.00544) & (0.00490) & (0.00482) \\
 & [0.941] & [0.866] & [0.485] & [0.577] \\
Centrality (std.) of i, Linked j,k & 0.00944 & 0.0102 & 0.00644 & 0.00424 \\
 & (0.00577) & (0.00565) & (0.00521) & (0.00523) \\
 & [0.106] & [0.074] & [0.220] & [0.420] \\
Avg. centrality (std.) of j,k, Not linked j,k & 0.00927 & 0.00697 & 0.00272 & 0.000500 \\
 & (0.00784) & (0.00758) & (0.00653) & (0.00663) \\
 & [0.241] & [0.361] & [0.679] & [0.940] \\
Avg. centrality (std.) of j,k, Linked j,k & -0.00227 & -0.00296 & -0.000628 & -0.00357 \\
 & (0.00602) & (0.00602) & (0.00369) & (0.00404) \\
 & [0.707] & [0.625] & [0.865] & [0.380] \\
 &  &  &  &  \\
Observations & 22,270 & 22,270 & 42,750 & 42,750 \\
Know pair only & $ $ & $ $ & \checkmark & \checkmark \\
Dimension FE & N/A & N/A & \checkmark & \checkmark \\
Demographic controls & $ $ & \checkmark & $ $ & \checkmark \\
 Depvar mean & 0.640 & 0.640 & 0.160 & 0.160 \\ \hline
\end{tabular}

%% file: bct_Feb19_2018.bbl
\begin{thebibliography}{31}
\newcommand{\enquote}[1]{``#1''}
\expandafter\ifx\csname natexlab\endcsname\relax\def\natexlab#1{#1}\fi

\bibitem[\protect\citeauthoryear{Acemoglu, Chernozhukov, and Yildiz}{Acemoglu
  et~al.}{2016}]{DaronFragility}
\textsc{Acemoglu, D., V.~Chernozhukov, and M.~Yildiz} (2016):
  \enquote{{Fragility of asymptotic agreement under Bayesian learning},}
  \emph{Theoretical Economics}, 11, 187--225.

\bibitem[\protect\citeauthoryear{Acemoglu, Dahleh, Lobel, and
  Ozdaglar}{Acemoglu et~al.}{2011}]{acemoglu2011bayesian}
\textsc{Acemoglu, D., M.~A. Dahleh, I.~Lobel, and A.~Ozdaglar} (2011):
  \enquote{Bayesian learning in social networks,} \emph{The Review of Economic
  Studies}, 78, 1201--1236.

\bibitem[\protect\citeauthoryear{Alatas, Banerjee, Chandrasekhar, Hanna, and
  Olken}{Alatas et~al.}{2016}]{alatas2012network}
\textsc{Alatas, V., A.~Banerjee, A.~G. Chandrasekhar, R.~Hanna, and B.~A.
  Olken} (2016): \enquote{Network structure and the aggregation of information:
  Theory and evidence from Indonesia,} \emph{American Economic Review}, 106,
  1663--1704.

\bibitem[\protect\citeauthoryear{Ballester, Calv{\'o}-Armengol, and
  Zenou}{Ballester et~al.}{2006}]{ballester2006s}
\textsc{Ballester, C., A.~Calv{\'o}-Armengol, and Y.~Zenou} (2006):
  \enquote{Who's who in networks. Wanted: The key player,} \emph{Econometrica},
  74, 1403--1417.

\bibitem[\protect\citeauthoryear{Banerjee, Chandrasekhar, Duflo, and
  Jackson}{Banerjee et~al.}{2013}]{banerjee2013diffusion}
\textsc{Banerjee, A., A.~G. Chandrasekhar, E.~Duflo, and M.~O. Jackson} (2013):
  \enquote{The diffusion of microfinance,} \emph{Science}, 341, 1236498.

\bibitem[\protect\citeauthoryear{Banerjee, Chandrasekhar, Duflo, and
  Jackson}{Banerjee et~al.}{2016}]{gossip2016}
---\hspace{-.1pt}---\hspace{-.1pt}--- (2016): \enquote{Gossip: Identifying
  Central Individuals in a Social Network,} NBER Working Paper No. 20422.

\bibitem[\protect\citeauthoryear{Banerjee}{Banerjee}{1992}]{banerjee1992simple}
\textsc{Banerjee, A.~V.} (1992): \enquote{A simple model of herd behavior,}
  \emph{The Quarterly Journal of Economics}, 107, 797--817.

\bibitem[\protect\citeauthoryear{Belloni and Chernozhukov}{Belloni and
  Chernozhukov}{2013}]{belloni2009least}
\textsc{Belloni, A. and V.~Chernozhukov} (2013): \enquote{Least squares after
  model selection in high-dimensional sparse models,} \emph{Bernoulli}, 19,
  521--547.

\bibitem[\protect\citeauthoryear{Belloni, Chernozhukov, and Hansen}{Belloni
  et~al.}{2014{\natexlab{a}}}]{belloni2014high}
\textsc{Belloni, A., V.~Chernozhukov, and C.~Hansen} (2014{\natexlab{a}}):
  \enquote{High-dimensional methods and inference on structural and treatment
  effects,} \emph{The Journal of Economic Perspectives}, 29--50.

\bibitem[\protect\citeauthoryear{Belloni, Chernozhukov, and Hansen}{Belloni
  et~al.}{2014{\natexlab{b}}}]{belloni2014inference}
---\hspace{-.1pt}---\hspace{-.1pt}--- (2014{\natexlab{b}}): \enquote{Inference
  on treatment effects after selection among high-dimensional controls,}
  \emph{The Review of Economic Studies}, 81, 608--650.

\bibitem[\protect\citeauthoryear{Bikhchandani, Hirshleifer, and
  Welch}{Bikhchandani et~al.}{1992}]{BHW}
\textsc{Bikhchandani, S., D.~Hirshleifer, and I.~Welch} (1992): \enquote{A
  theory of fads, fashion, custom, and cultural change as information
  cascades,} \emph{Journal of Political Economy}, 100, 992--1026.

\bibitem[\protect\citeauthoryear{Calv\'{o}-Armengol, Patacchini, and
  Zenou}{Calv\'{o}-Armengol et~al.}{2009}]{CalvoREStud}
\textsc{Calv\'{o}-Armengol, A., E.~Patacchini, and Y.~Zenou} (2009):
  \enquote{Peer effects and social networks in education,} \emph{Review of
  Economic Studies}, 76, 1239--1267.

\bibitem[\protect\citeauthoryear{Candogan, Bimpikis, and Ozdaglar}{Candogan
  et~al.}{2012}]{Ozan}
\textsc{Candogan, O., K.~Bimpikis, and A.~Ozdaglar} (2012): \enquote{Optimal
  pricing in networks with externalities,} \emph{Operations Research}, 60,
  883--905.

\bibitem[\protect\citeauthoryear{de~Paula, Richards-Shubik, and Tamer}{de~Paula
  et~al.}{2018}]{de2017identifying}
\textsc{de~Paula, A., S.~Richards-Shubik, and E.~Tamer} (2018):
  \enquote{Identifying preferences in networks with bounded degree,}
  \emph{Econometrica}, 86, 263--288.

\bibitem[\protect\citeauthoryear{DeGroot}{DeGroot}{1974}]{degroot1974reaching}
\textsc{DeGroot, M.~H.} (1974): \enquote{Reaching a consensus,} \emph{Journal
  of the American Statistical Association}, 69, 118--121.

\bibitem[\protect\citeauthoryear{Eyster and Rabin}{Eyster and
  Rabin}{2014}]{EysterRabinQJE}
\textsc{Eyster, E. and M.~Rabin} (2014): \enquote{Extensive imitation is
  irrational and harmful,} \emph{The Quarterly Journal of Economics}, 129,
  1861--1898.

\bibitem[\protect\citeauthoryear{Friedkin}{Friedkin}{1983}]{friedkin1983horizons}
\textsc{Friedkin, N.~E.} (1983): \enquote{Horizons of observability and limits
  of informal control in organizations,} \emph{Social Forces}, 62, 54--77.

\bibitem[\protect\citeauthoryear{Goyal and Moraga-Gonz\'{a}lez}{Goyal and
  Moraga-Gonz\'{a}lez}{2001}]{GoyalRAND}
\textsc{Goyal, S. and J.~L. Moraga-Gonz\'{a}lez} (2001): \enquote{R\&\textsc{D}
  networks,} \emph{The RAND Journal of Economics}, 32, 686--707.

\bibitem[\protect\citeauthoryear{Krackhardt}{Krackhardt}{1987}]{krackhardt1987cognitive}
\textsc{Krackhardt, D.} (1987): \enquote{Cognitive social structures,}
  \emph{Social networks}, 9, 109--134.

\bibitem[\protect\citeauthoryear{Krackhardt}{Krackhardt}{2014}]{krackhardt2014preliminary}
---\hspace{-.1pt}---\hspace{-.1pt}--- (2014): \enquote{A preliminary look at
  accuracy in egonets,} in \emph{Contemporary Perspectives on Organizational
  Social Networks}, Emerald Group Publishing Limited, 277--293.

\bibitem[\protect\citeauthoryear{Leung}{Leung}{2015{\natexlab{a}}}]{leung2014random}
\textsc{Leung, M.~P.} (2015{\natexlab{a}}): \enquote{A random-field approach to
  inference in large models of network formation,} \emph{Available at SSRN}.

\bibitem[\protect\citeauthoryear{Leung}{Leung}{2015{\natexlab{b}}}]{leung2015two}
---\hspace{-.1pt}---\hspace{-.1pt}--- (2015{\natexlab{b}}): \enquote{Two-step
  estimation of network-formation models with incomplete information,}
  \emph{Journal of Econometrics}, 188, 182--195.

\bibitem[\protect\citeauthoryear{Li and Tan}{Li and Tan}{2017}]{li2017learning}
\textsc{Li, W. and X.~Tan} (2017): \enquote{Learning in local networks,}
  Working paper.

\bibitem[\protect\citeauthoryear{Lobel and Sadler}{Lobel and
  Sadler}{2015}]{LobelSadlerTE}
\textsc{Lobel, I. and E.~Sadler} (2015): \enquote{Information diffusion in
  networks through social learning,} \emph{Theoretical Economics}, 10,
  807--851.

\bibitem[\protect\citeauthoryear{Mele}{Mele}{2017}]{Mele2017}
\textsc{Mele, A.} (2017): \enquote{A structural model of dense network
  formation,} \emph{Econometrica}, 85, 825--850.

\bibitem[\protect\citeauthoryear{Menzel}{Menzel}{2017}]{menzel2015strategic}
\textsc{Menzel, K.} (2017): \enquote{Strategic network formation with many
  agents,} Working paper.

\bibitem[\protect\citeauthoryear{Mossel, Olsman, and Tamuz}{Mossel
  et~al.}{2016}]{MosselOlsmanTamuz}
\textsc{Mossel, E., N.~Olsman, and O.~Tamuz} (2016): \enquote{{Efficient
  Bayesian learning in social networks with Gaussian estimators},} in
  \emph{Annual Allerton Conference on Communication, Control, and Computing},
  Allerton House, Illinois, 425--432.

\bibitem[\protect\citeauthoryear{Mossel, Sly, and Tamuz}{Mossel
  et~al.}{2015}]{mossel2015strategic}
\textsc{Mossel, E., A.~Sly, and O.~Tamuz} (2015): \enquote{Strategic learning
  and the topology of social networks,} \emph{Econometrica}, 83, 1755--1794.

\bibitem[\protect\citeauthoryear{Ridder and Sheng}{Ridder and
  Sheng}{2017}]{RidderSheng}
\textsc{Ridder, G. and S.~Sheng} (2017): \enquote{Estimation of large network
  formation games,} Working paper.

\bibitem[\protect\citeauthoryear{Sheng}{Sheng}{2016}]{sheng2016structural}
\textsc{Sheng, S.} (2016): \enquote{A structural econometric analysis of
  network formation games,} Working paper.

\bibitem[\protect\citeauthoryear{Smith and S{\o}rensen}{Smith and
  S{\o}rensen}{2000}]{SmithSorensen}
\textsc{Smith, L. and P.~N. S{\o}rensen} (2000): \enquote{Pathological outcomes
  of observational learning,} \emph{Econometrica}, 68, 371--398.

\end{thebibliography}
